**Superfluid Brillouin Optomechanics**

A. D. Kashkanova,[1] A. B. Shkarin,[1] C. D. Brown,[1] N. E. Flowers-Jacobs,[1] L. Childress,[1,2] S. W. Hoch,[1] L. Hohmann,[3] K. Ott,[3] J. Reichel,[3] and J. G. E. Harris[1,4]

[1] *Department of Physics, Yale University, New Haven, CT, 06511, USA*
[2] *Department of Physics, McGill University, 3600 Rue University, Montreal, Quebec H3A 2T8, Canada*
[3] *Laboratoire Kastler Brossel, ENS-PSL Research University, CNRS, UPMC-Sorbonne Universités, Collège de France, F-75005 Paris, France*
[4] *Department of Applied Physics, Yale University, New Haven, CT, 06511, USA*

Optomechanical systems couple an electromagnetic cavity to a mechanical resonator which is typically formed from a solid object. The range of phenomena accessible to these systems depends on the properties of the mechanical resonator and on the manner in which it couples to the cavity fields. In both respects, a mechanical resonator formed from superfluid liquid helium offers several appealing features: low electromagnetic absorption, high thermal conductivity, vanishing viscosity, well-understood mechanical loss, and *in situ* alignment with cryogenic cavities. In addition, it offers degrees of freedom that differ qualitatively from those of a solid. Here, we describe an optomechanical system consisting of a miniature optical cavity filled with superfluid helium. The cavity mirrors define optical and mechanical modes with near-perfect overlap, resulting in an optomechanical coupling rate ~ 3 kHz. This coupling is used to drive the superfluid; it is also used to observe the superfluid's thermal motion, resolving a mean phonon number as low as 11.

Light confined in a cavity exerts forces on the components that form the cavity. These forces can excite mechanical vibrations in the cavity components, and these vibrations can alter the propagation of light in the cavity. This interplay between electromagnetic (EM) and mechanical degrees of freedom is the basis of cavity optomechanics. It gives rise to a variety of nonlinear phenomena in both the EM and mechanical domains, and provides means for controlling and sensing EM fields and mechanical oscillators.[1]

If the optomechanical interaction is approximately unitary, it can provide access to quantum effects in the optical and mechanical degrees of freedom.[1] Optomechanical systems have been used to observe quantum effects which are remarkable in that they are associated with the motion of massive objects.[2,3,4,5,6,7,8,9,10,11,12] They have also been proposed for use in a range of quantum information and sensing applications.[13,14,15,16,17,18,19,20,21,22] Realizing these goals typically requires strong optomechanical coupling, weak EM and mechanical loss, efficient cooling to cryogenic temperatures, and reduced influence from technical noise.

To date, nearly all optomechanical devices have used solid objects as mechanical oscillators. However, liquid oscillators offer potential advantages. A liquid can conformally fill a hollow EM cavity,[23] allowing for near-perfect overlap between the cavity's EM modes and the normal modes of the liquid body's vibrations. In addition, the liquid's composition can be changed *in situ*, an important feature for applications in fluidic sensing.[24] However, most liquids face two important obstacles to operation in or near the quantum regime: their viscosity results in strong mechanical losses, and they solidify when cooled to cryogenic temperatures. Liquid helium is exceptional in both respects, as it does not solidify under its own vapor pressure and possesses zero viscosity in its purely superfluid state. In addition, liquid He has low EM loss and high thermal conductivity at cryogenic temperatures.

The interaction of light with the mechanical modes of superfluid He has been studied in a variety of contexts, including free-space spontaneous inelastic light scattering (i.e., without a cavity) from thermal excitations of first sound, second sound, isotopic concentration, ripplons and rotons.[25,26,27,28,29] These experiments were carried out at relatively high temperatures ($T \sim 1-2$ K) where these excitations are strongly damped. The combination of strong damping and the lack of confinement for the optical or mechanical modes precluded observation of cavity optomechanical behavior. More recently, cavity optomechanical interactions were measured between near infrared (NIR) light in a cavity and the third sound modes of a superfluid He film coating the cavity; this interaction was used to monitor and control the superfluid's thermal motion, but was predominantly of non-unitary photothermal origin (and therefore not well-suited to studying quantum optomechanical effects).[30] In the microwave domain, a cavity was used to monitor the externally-driven acoustic (first sound) modes of superfluid He inside the cavity.[31] This device demonstrated very high acoustic quality factor (~$10^7$) and EM quality factor (~$10^7$); however the weak optomechanical coupling (estimated single-photon coupling rate ~ $4 \times 10^{-8}$ Hz) precluded observation of the superfluid's thermal motion or the cavity field's influence upon the acoustic modes.

Here we describe an optomechanical system consisting of a NIR optical cavity filled with superfluid He. We observe coupling between the cavity's optical modes and the superfluid's acoustic modes, and find that this coupling is predominantly electrostrictive in origin (and hence unitary). The single-photon coupling rate is ~ $3 \times 10^3$ Hz, enabling observation of the superfluid's thermal motion and of the cavity field's influence upon the acoustic modes. These modes are cooled to 180 mK (corresponding to mean phonon number 11), and reach a maximum quality factor $6 \times 10^4$. These results agree with a simple model using well-known material properties, and may be improved substantially via straightforward modifications of the present device.

The system is shown schematically in Fig. 1(a). The optical cavity is formed between a pair of single-mode optical fibers. Laser machining is used to produce a smooth concave surface on the face of each fiber.[32] The radii of curvature of the two faces are $r_1 = 409$ μm and $r_2 = 282$ μm. On each face, alternating layers of $SiO_2$ and $Ta_2O_5$ are deposited to form a distributed Bragg reflector (DBR).[33] The DBRs' power transmissions are $T_1 = 1.03 \times 10^{-4}$ and $T_2 = 1.0 \times 10^{-5}$; as a result the cavity is approximately single-sided. The fibers are aligned in a glass ferrule with inner diameter $133 \pm 5$ μm, as described in Ref. [34]. The ferrule is epoxied into a brass cell, and the fibers are fixed so that the cavity length $L = 67.3$ μm after cooling to cryogenic temperatures. The cell is mounted on the mixing chamber (MC) of a dilution refrigerator. Liquid He is introduced into the cell via a fill line.

A closer view of the cavity is shown in the lower portion of Fig. 1(a). The optical modes are confined by the DBRs' high reflectivity and concave shape. The acoustic modes are confined in the same manner (the acoustic reflectivity is primarily due to impedance mismatch between He and the DBR materials, see the Supplementary Information).

Coupling between optical and acoustic modes arises because the spatial variation of He density associated with an acoustic mode can alter the effective cavity length for an optical mode; equivalently, the intensity variation associated with an optical mode can exert an electrostrictive force that drives an acoustic mode. This coupling is described by the conventional optomechanical

Hamiltonian[1] $\hat{H}_{OM} = \hbar g_0^{\alpha,\beta} \hat{a}_\alpha^\dagger \hat{a}_\alpha (\hat{b}_\beta^\dagger + \hat{b}_\beta)$ where $\hat{a}_\alpha^\dagger$ is the photon creation operator for the optical mode $\alpha$, $\hat{b}_\beta^\dagger$ is the phonon creation operator for the acoustic mode $\beta$, and the single-photon coupling rate is

$$g_0^{\alpha,\beta} = \omega_\alpha (n_{He} - 1) \int \tilde{\rho}_{\beta,zp} B_\beta(\vec{r}) |\vec{A}_\alpha(\vec{r})|^2 d^3\vec{r} \qquad (1)$$

Here $\vec{A}_\alpha(\vec{r})$ and $B_\beta(\vec{r})$ are dimensionless, square-normalized functions representing the electric field of the optical mode $\alpha$ and the density variation of the acoustic mode $\beta$ respectively. $\tilde{\rho}_{\beta,zp}$ is the fractional density change associated with the zero-point fluctuations of the acoustic mode $\beta$, and is defined by $K_{He} \int (\tilde{\rho}_{\beta,zp} B_\beta(\vec{r}))^2 d^3\vec{r} = \hbar \omega_\beta / 2$. The frequency of the optical (acoustic) mode is $\omega_\alpha$ ($\omega_\beta$). The bulk modulus and index of refraction of liquid He are $K_{He} = 8.21 \times 10^6$ Pa and $n_{He} = 1.028$.

The normal modes $\vec{A}_\alpha(\vec{r})$ and $B_\beta(\vec{r})$ can be found by noting that, inside the cavity, Maxwell's equations and the hydrodynamic equations both reduce to wave equations: the former owing to the absence of EM sources, and the latter under the assumption that the liquid undergoes irrotational flow with small velocity, small displacements, and small variations in pressure and density.[35] Furthermore, the cavity geometry (set by $r_1$, $r_2$, and $L$) and the wavelengths of interest (discussed below) allow the paraxial approximation to be used, leading to well-known solutions.[36] As a result, the optical and acoustic modes have the same general form: a standing wave along the cavity axis and a transverse profile described by two Gaussian Hermite polynomials. Each mode is indexed by three positive integers, i.e.: $\alpha = \{x_\alpha, y_\alpha, z_\alpha\}$ and $\beta = \{x_\beta, y_\beta, z_\beta\}$ where the $x_\alpha, y_\alpha$ ($x_\beta, y_\beta$) index the transverse optical (acoustic) mode, and $z_\alpha$ ($z_\beta$) is the number of optical (acoustic) half-wavelengths in the cavity. Here, we consider only the lowest-order transverse modes (i.e., those with $x_\alpha, y_\alpha, x_\beta, y_\beta = 0$).

Boundary conditions ensure that the interface between the DBR and the liquid He corresponds (nearly) to a node of $A_\alpha$ and an antinode of $B_\beta$. As a result, it is straightforward to show that $g_0^{\alpha,\beta}$ nearly vanishes unless $2z_\alpha = z_\beta$. This requirement is equivalent to the phase-matching condition for stimulated Brillouin scattering[37] applied to the standing waves of a paraxial cavity. Thus, an optical mode with wavelength (in liquid He) $\lambda_\alpha$ = 1.50 μm ($\omega_\alpha/2\pi$ = 195 THz) will couple primarily to a single acoustic mode with wavelength $\lambda_\beta$ = 0.75 μm ($\omega_\beta/2\pi$ = 318 MHz). A derivation of these features from first principles is given in Ref.[38].

The measurement setup is illustrated in Fig 1(b),(c). Light from a tunable laser (1,520 nm < $\lambda$ < 1,560 nm) passes through a frequency shifter (FS) and a phase modulator (PM). The PM is driven by as many as three tones, resulting in first-order sidebands labeled as lock, probe, and control in Fig. 1(c), while the carrier beam serves as a local oscillator (LO). Light is delivered to (and collected from) the cryostat via a circulator. Light leaving the cavity passes through an erbium-doped fiber amplifier (EDFA) and then is detected by a photodiode.

The lock beam is produced by a fixed frequency drive ($\omega_{lock}/2\pi$ = 926 MHz). The beat note between the reflected lock beam and LO beam produces an error signal that is used to control the FS, ensuring that all the beams track fluctuations in the cavity.

The probe beam is produced by the variable frequency drive ($\omega_{probe}$) from a vector network analyzer (VNA). The beat note between the reflected probe and LO beams is monitored by the VNA. The red data in Fig. 1(d) shows the intracavity power inferred from the VNA signal as $\omega_{probe}$ is varied to scan the probe beam over the optical mode with $z_\alpha$ = 90. This data is taken without He in the cell and with the refrigerator temperature $T_{MC}$ = 30 mK. Fitting this data gives the decay rate $\kappa_\alpha/2\pi$ = 46.1 ± 0.1 MHz, typical of the decay rates measured with the cavity at room temperature.

To determine whether the presence of liquid He alters the cavity's optical loss, the blue data in Fig. 1(d) shows the same measurement after the cavity is filled with liquid He. For this measurement, $z_\alpha$ = 93 and $T_{MC}$ = 38 mK. Fitting this data gives $\kappa_\alpha/2\pi$ = 46.3 ± 0.2 MHz. The difference between the two values of $\kappa_\alpha$ is consistent with the variations between modes when the cavity is empty, and is also consistent with the negligible optical loss expected for liquid He at these temperatures.[39]

The acoustic modes of the liquid He were characterized via optomechanically induced amplification (OMIA).[40] To accomplish this, a control beam was produced by the variable frequency ($\omega_{control}$) drive of a microwave generator. When the difference $|\omega_{control} - \omega_{probe}| \approx \omega_\beta$, the intracavity beating between the control and probe beams can excite the acoustic mode $\beta$; the resulting acoustic oscillations modulate the control beam, and these modulations are detected by the VNA (see Supplementary Information).

Fig. 2(a) shows a typical record of the normalized amplitude $a$ and phase $\psi$ of the VNA signal when the control beam is detuned from the cavity resonance by $\Delta_{control} \approx \omega_\beta$ and $\omega_{probe}$ is varied. The peak at $|\omega_{control} - \omega_{probe}|/2\pi \approx 317.32$ MHz corresponds to the resonance of the acoustic mode. The solid line in Fig. 2(a) is a fit to the expected form of $a$ and $\psi$ (see the Supplementary Information). The fit parameters are $\omega_\beta$, $\gamma_\beta$ (the acoustic damping rate), as well as $A$ and $\Psi$ (the overall amplitude and phase of the OMIA lineshape, described in Supplementary Information),.

Figure 2(b) shows $A$ and $\Psi$ (extracted from fits similar to the one in Fig. 2(a)) as a function of $\Delta_{control}$ and $P_{control}$ (the power of the control beam). For each $P_{control}$, $A$ shows a peak of width $\kappa_\alpha/2\pi$ centered at $\Delta_{control} = \omega_\beta$, while $\Psi$ changes by $\sim \pi$ over the same range of $\Delta_{control}$. These features correspond to the excitation of the optical resonance by the probe beam. Fitting the measurements of $A$ and $\Psi$ to the expected form of the OMIA response (assuming purely electrostrictive coupling) gives the dashed lines in Fig. 2(b). This fit has only one parameter ($g_0^{\alpha,\beta}$), and returns the best-fit value $g_0^{\alpha,\beta}/2\pi = 3.3 \pm 0.2 \times 10^3$ Hz; in comparison, numerical evaluation of Eq. 1 gives $g_0^{\alpha,\beta}/2\pi = 3.5 \pm 0.5 \times 10^3$ Hz.

While this fit captures many features of the data, it underestimates $\Psi$ by an amount roughly independent of $\Delta_{control}$ and $P_{control}$. To account for this constant phase shift, we calculated the OMIA signal that would result if the optical intensity also drives the acoustic mode via an interaction mediated by a process much slower than $\omega_\beta$ (see the Supplementary Information). Such a slow interaction would arise naturally from a photothermal process in which optical absorption in the DBR heats the liquid He. The solid lines in Fig. 2(b) are a fit to this calculation, in which the fit parameters are $g_0^{\alpha,\beta}$ and $g_{0,pt}^{\alpha,\beta}$ (the single-photon photothermal coupling rate), and the best-fit values are $g_0^{\alpha,\beta}/2\pi = 3.18 \pm 0.2 \times 10^3$ Hz and $g_{0,pt}^{\alpha,\beta}/2\pi = 0.97 \pm 0.05 \times 10^3$ Hz respectively.

The damping of the acoustic modes is expected to be dominated by two processes. The first is mode conversion via the nonlinear compressibility of liquid He. This process has been studied extensively,[41] and for the relevant temperature range would result in an acoustic quality factor $Q_{\beta,int} = \chi / T_{bath}^4$, where $\chi = 118$ K$^4$ (see the Supplementary Information) and $T_{bath}$ is the temperature of the He in the cavity. The second expected source of damping is acoustic radiation from the He into the confining materials. This process is predicted to result in $Q_{\beta,ext} = (79 \pm 5) \times 10^3$ (see the Supplementary Information).

Figure 3(a) shows $Q_\beta$ (determined from data and fits similar to Fig. 2(a)) as a function of $T_{MC}$ and $P_{inc}$ (the total laser power incident on the cavity). Also shown are the predicted $Q_{\beta,int}$ and $Q_{\beta,ext}$

(dashed lines), and their combined effect assuming $T_{\text{bath}} = T_{\text{MC}}$ (black line). Although the data show qualitative agreement with the predicted trends, there is also a clear dependence of $Q_\beta$ upon $P_{\text{inc}}$. Figure 3(b) shows that $Q_\beta$ depends on the mean intracavity photon number $\bar{n}_\alpha$ as well as $P_{\text{inc}}$.

For the conditions of these measurements, dynamical backaction[1] (i.e., optical damping) is not expected to contribute appreciably to $Q_\beta$. Instead, we consider a model in which light is absorbed in the DBRs, resulting in heat flow into the cavity $\Phi = \mu P_{\text{inc}} + \nu \hbar \omega_\alpha \kappa_{\alpha,\text{int}} \bar{n}_\alpha$. Here $\kappa_{\alpha,\text{int}}$ is the intracavity loss rate (and is determined from measurements similar to those shown in Fig. 1(d)), while $\nu$ and $\mu$ are dimensionless constants that characterize, respectively, the absorbers' overlap with the optical standing mode (which extends into the upper layers of the DBRs) and the optical travelling mode. In the presence of $\Phi$, equilibrium is maintained by the thermal conductance between the cavity and the MC, which is dominated by the sheath of liquid He between the optical fibers and the glass ferrule (Fig. 1(a)). For the relevant temperature range, the conductance of liquid He is $k = \varepsilon T^3$, where $T$ is the local temperature and $\varepsilon$ is measured to be $(5 \pm 2.5) \times 10^{-5}$ W/K$^4$ (see the Supplementary Information). This model predicts that

$$T_{\text{bath}} = (T_{\text{MC}}^4 + \tfrac{4}{\varepsilon}(\mu P_{\text{inc}} + \nu \hbar \omega_\alpha \kappa_{\alpha,\text{int}} \bar{n}_\alpha))^{1/4} . \tag{2}$$

The colored solid lines in Fig 3(a) and 3(b) are the result of fitting the complete dataset to $Q_\beta(T_{\text{bath}}) = (Q_{\beta,\text{int}}^{-1}(T_{\text{bath}}) + Q_{\beta,\text{ext}}^{-1})^{-1}$ by using Eq. 2 and taking $\nu/\mu$, $\mu/\varepsilon$, and $Q_{\beta,\text{ext}}$ as fitting parameters. This fit gives $\nu/\mu = 294 \pm 9$, $\mu/\varepsilon = 48 \pm 7$ K$^4$/W, and $Q_{\beta,\text{ext}} = (70 \pm 2.0) \times 10^3$. The value of $\nu/\mu$ is consistent with absorbers being distributed throughout the DBR layers, and the value of $Q_{\beta,\text{ext}}$ is consistent with the *a priori* calculation in the Supplementary Information.

Figure 3(c) shows the data from Figs. 3(a) and 3(b) replotted as a function of $T_{\text{bath}}$ (calculated from Eq. 2 and the best-fit values of $\nu/\mu$ and $\mu/\varepsilon$). The data collapse together, indicating that $Q_\beta$ is determined by $T_{\text{bath}}$, which in turn is determined by $T_{\text{MC}}$, $P_{\text{inc}}$, and $\bar{n}_\alpha$ in accordance with the model described above. The collapsed data are in close agreement with the prediction for $Q_\beta(T_{\text{bath}})$ (the black line in Fig. 3(c)).

To determine $\bar{n}_\beta$ (the mean phonon number of the acoustic mode), a heterodyne technique was used to measure the Stokes sideband imprinted on the control beam by the acoustic mode's thermal fluctuations. For these measurements, the probe beam was turned off and the control beam's detuning relative to the cavity resonance was set to $\Delta_{\text{control}} \approx \omega_\beta$. A spectrum analyzer was used to monitor the photocurrent at frequencies near the beat note between the Stokes sideband and the LO. Fig. 4(a) shows $S_{II}(\omega)$, the power spectral density (PSD) of the photocurrent, for a range of $T_{\text{MC}}$ and $P_{\text{inc}}$. As $T_{\text{MC}}$ and $P_{\text{inc}}$ are reduced the acoustic resonances become narrower, in qualitative agreement with Fig. 3(a) and 3(b).

Figure 4(b) shows the same data as in Fig. 4(a), but with $S_{II}$ converted to $S_{\tilde{\rho}\tilde{\rho}}$ (the PSD of fractional density fluctuations), and normalized by $4\tilde{\rho}^2_{\beta,\text{zp}}/\gamma_\beta$ (see Supplemental Information) so that the peak height corresponds to $\bar{n}_\beta$. The solid lines are fits to the expected Lorentzian lineshape. Fig. 4(c) shows $\bar{n}_\beta$ determined from these fits and plotted as a function of $T_{\text{bath}}$ (which is determined using Eq. 2). Also shown is the solid red line corresponding to the prediction $\bar{n}_\beta = k_B T_{\text{bath}}/\hbar\omega_\beta$. The data and prediction show close agreement for $\bar{n}_\beta$ as low as $11 \pm 0.3$, indicating that the acoustic mode remains in thermal equilibrium with the material temperature $T_{\text{bath}}$.

In conclusion, these results show a promising combination of cavity optomechanics with a superfluid. This system achieves dimensionless figures of merit comparable to state-of-the-art, solid-based optomechanical systems ($\omega_\beta/\kappa_\alpha = 4.6$, $g_0^{\alpha,\beta}/\kappa_\alpha = 4.6 \times 10^{-5}$, $k_B T_{\text{bath}}/\hbar\omega_\beta = 11$), and without the need for *in situ* alignment. The acoustic loss in this system agrees well with a simple model, and this model indicates that straightforward refinements can provide substantial improvements. For example, $Q_\beta$ may be increased to $\sim 3 \times 10^6$ by using DBR structures that serve as high-reflectivity acoustic mirrors as well as optical mirrors (see Supplementary Information). In addition, $T_{\text{bath}}$ can be lowered by increasing the thermal conductance between the cavity and the refrigerator (e.g., using the device geometry described in Ref. [34]).

These results also open a number of qualitatively new directions, including cavity optomechanical coupling to degrees of freedom that are unique to superfluid liquid He. These include the Kelvin modes of remnant vortex lines; ripplon modes of the superfluid's free surface; and well-controlled impurities, such as electrons on the surface or in the bulk of the superfluid. Precision measurements of these degrees of freedom may provide new insight into long-standing questions about their roles in superfluid turbulence[42,43] and their potential applications in quantum information processing.[44]

**Acknowledgements**

We are grateful to Vincent Bernardo, Joe Chadwick, John Cummings, Andreas Fragner, Katherine Lawrence, Donghun Lee, Daniel McKinsey, Peter Rakich, Robert Schoelkopf, Hong Tang, Jedidiah Thompson, and Zuyu Zhao for their assistance. We acknowledge financial support from W. M. Keck Foundation Grant No. DT121914, AFOSR Grants FA9550-09-1-0484 and FA9550-15-1-0270, DARPA Grant W911NF-14-1-0354, ARO Grant W911NF-13-1-0104, and NSF Grant 1205861. This work has been supported by the DARPA/MTO ORCHID program through a grant from AFOSR. This project was made possible through the support of a grant from the John Templeton Foundation. The opinions expressed in this publication are those of the authors and do not necessarily reflect the views of the John Templeton Foundation. This material is based upon work supported by the National Science Foundation Graduate Research Fellowship under Grant No. DGE-1122492. L.H., K.O. and J.R. acknowledge funding from the EU Information and Communication Technologies program (QIBEC project, GA 284584), ERC (EQUEMI project, GA 671133), and IFRAF.

**Author Contributions**

A.K., A.S., and C.D.B. performed the measurements and analysis; A.K., A.S., and N.E.F.J. assembled the device; A.K. and L.C. built and tested prototypes of the device; S.W.H., L.H., and K.O. carried out the laser machining of the fibers; J.R. supervised the laser machining; J.G.E.H supervised the other phases of the project.

**Additional Information**

Supplementary information is available in the online version of the paper. Correspondence and requests for materials should be addressed to J.G.E.H.

**Competing Financial Interests**

The authors declare no competing financial interests.

**Figure Captions**

**Figure 1 | Description and characterization of the superfluid-filled optical cavity. a**, Schematic illustration of the device, showing the optical cavity formed between two optical fibers (yellow) aligned in a glass ferrule (yellow). The ferrule is mounted in a brass cell (grey), which is attached to a dilution refrigerator (not shown) and can be filled with superfluid helium (blue). The lower panel illustrates the optomechanical coupling: the intensity maxima of an optical mode (red line) overlap with the density maxima of an acoustic mode (blue shading; darker corresponds to denser He). **b**, Schematic of the measurement set up. Red: optical components. Green: electronic components. Light from a tunable laser (TL) passes through a frequency shifter (FS) and a phase modulator (PM). The PM is driven by tones from a voltage-controlled oscillator (VCO1), microwave generator (MG), and vector network analyzer (VNA). The resulting sidebands and the carrier are delivered to the cryostat by a circulator (circle), which also sends the reflected beams through an erbium-doped fiber amplifier (triangle) to a photodiode (PD). The photocurrent can be monitored by the VNA or a microwave spectrum analyzer (MSA). It is also mixed with the tone from VCO1 to produce an error signal that is sent to VCO2, which in turn drives the FS in order to lock the beams to the cavity. **c**, Illustration of the laser beams. The LO, control, lock, and probe beams are shown (red), along with the cavity lineshape (black). **d**, The intracavity power as the probe beam is detuned. Red: empty cavity. Blue: cavity filled with liquid He. To aid comparison, the data are normalized to their maximum value and plotted as a function of the detuning from the cavity resonance. The solid red and blue lines are fits to the expected Lorentzian, and give linewidths 46.1 ± 0.1 MHz and 46.3 ± 0.2 MHz, respectively.

**Figure 2 | Characterizing the acoustic mode and the optomechanical coupling. a**, The relative amplitude $a$ and phase $\psi$ of the OMIA signal as a function of the intracavity beat note frequency $|\omega_{control} - \omega_{probe}|/2\pi$. For these measurements $L = 85.2$ µm, $z_\alpha = 112$, $z_\beta = 224$, $\kappa_\alpha/2\pi = 69 \pm 2$ MHz, $\Delta_{control}/2\pi = 320$ MHz, $P_{control} = 4$ µW, and $T_{MC} = 59$ mK. The solid line is the fit described in the text and Supplementary Information. **b**, The amplitude $A$ and phase $\Psi$ of the OMIA lineshape (determined from fits similar to the one in **a**) as a function of $\Delta_{control}$ and $P_{control}$. The dashed line is the fit assuming only electrostrictive coupling; the solid line is the fit assuming electrostrictive and

slow photothermal coupling, as described in the text and Supplementary Information.

**Figure 3 | Acoustic damping as a function of temperature. a**, Acoustic quality factor $Q_\beta$ versus $T_{MC}$. Different colors correspond to different values of $P_{inc}$, the total optical power incident on the cavity. The red and blue dashed lines are the predicted contributions to $Q_\beta$ from the internal loss and radiation loss, respectively, assuming $T_{bath} = T_{MC}$; the solid black line is the net effect of these contributions. The colored solid lines are the fit to the model described in the text. **b**, $Q_\beta$ versus $\bar{n}_\alpha$ (the mean number of intracavity photons). Different colors correspond to different values of $P_{inc}$. The colored solid lines are the fit to the model described in the text. **c**, The data from **a** and **b**, plotted versus $T_{bath}$ (which is inferred from the fits in **a** and **b**). Different colors correspond to different values of $\bar{n}_\alpha$. The solid black line, as well as the dashed red and blue lines, are the same as in **a**.

**Figure 4 | Thermal fluctuations of the acoustic mode. a**, The power spectral density of the photocurrent for various $T_{MC}$ and $P_{inc}$ when no drive is applied to the acoustic mode. The measurement noise floor changes with $P_{inc}$ owing to the EDFA's nonlinearity. The solid lines are fits to a Lorentzian plus a constant background. **b**, The same data as in **a**, converted from photocurrent to helium density fluctuations. The background has been subtracted and the data normalized so that the peak height corresponds to the mean phonon number $\bar{n}_\beta$. **c**, The mean phonon number versus $T_{bath}$.

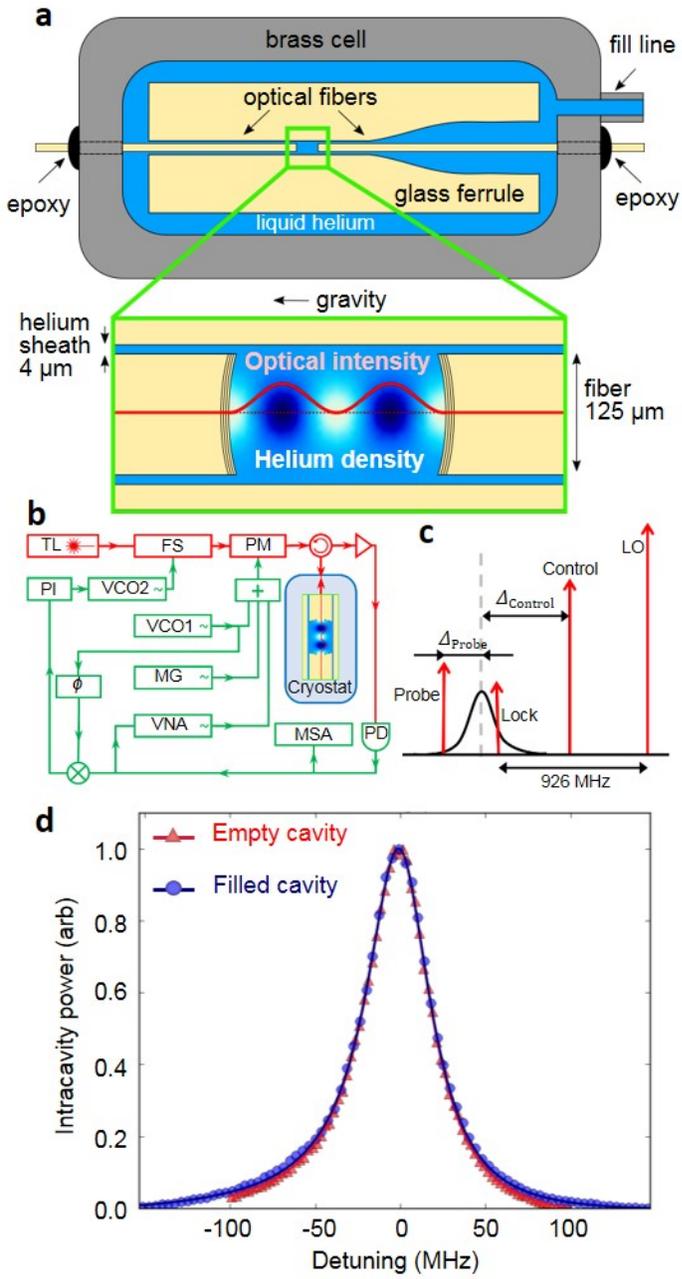

**Figure 1**

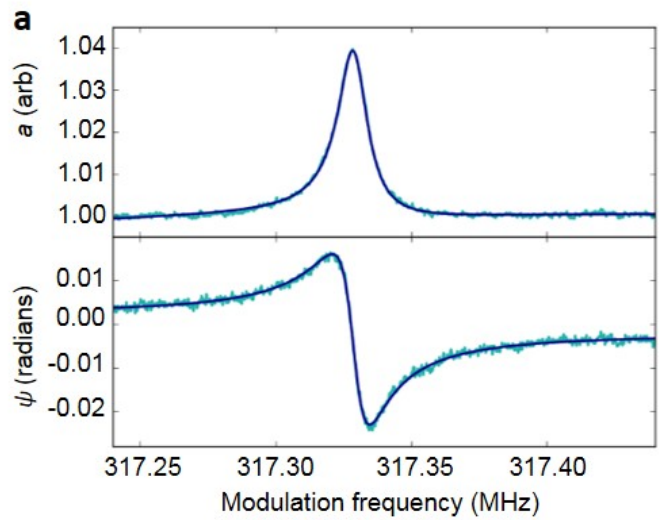
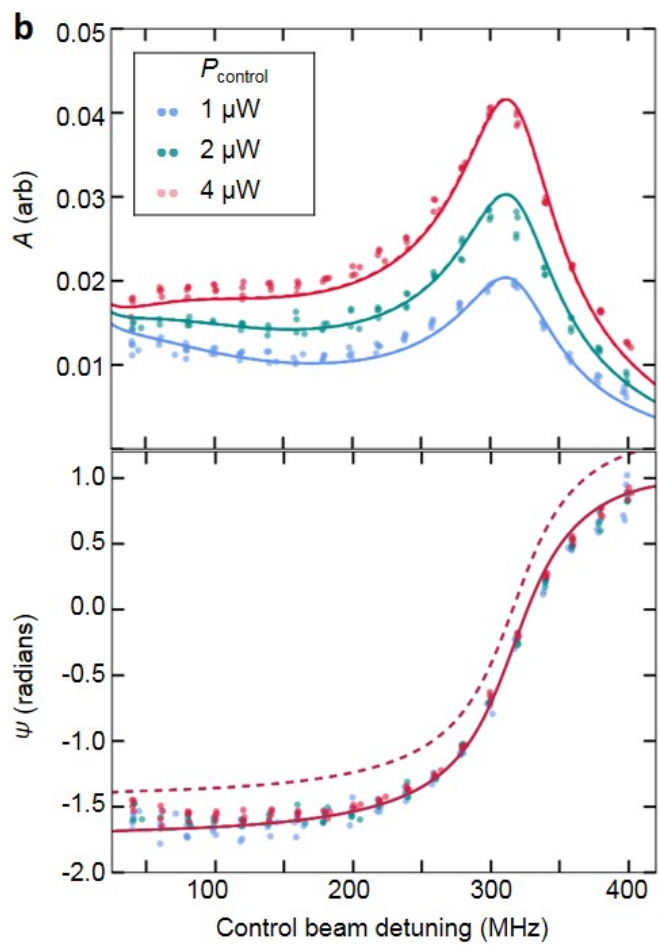

**Figure 2**

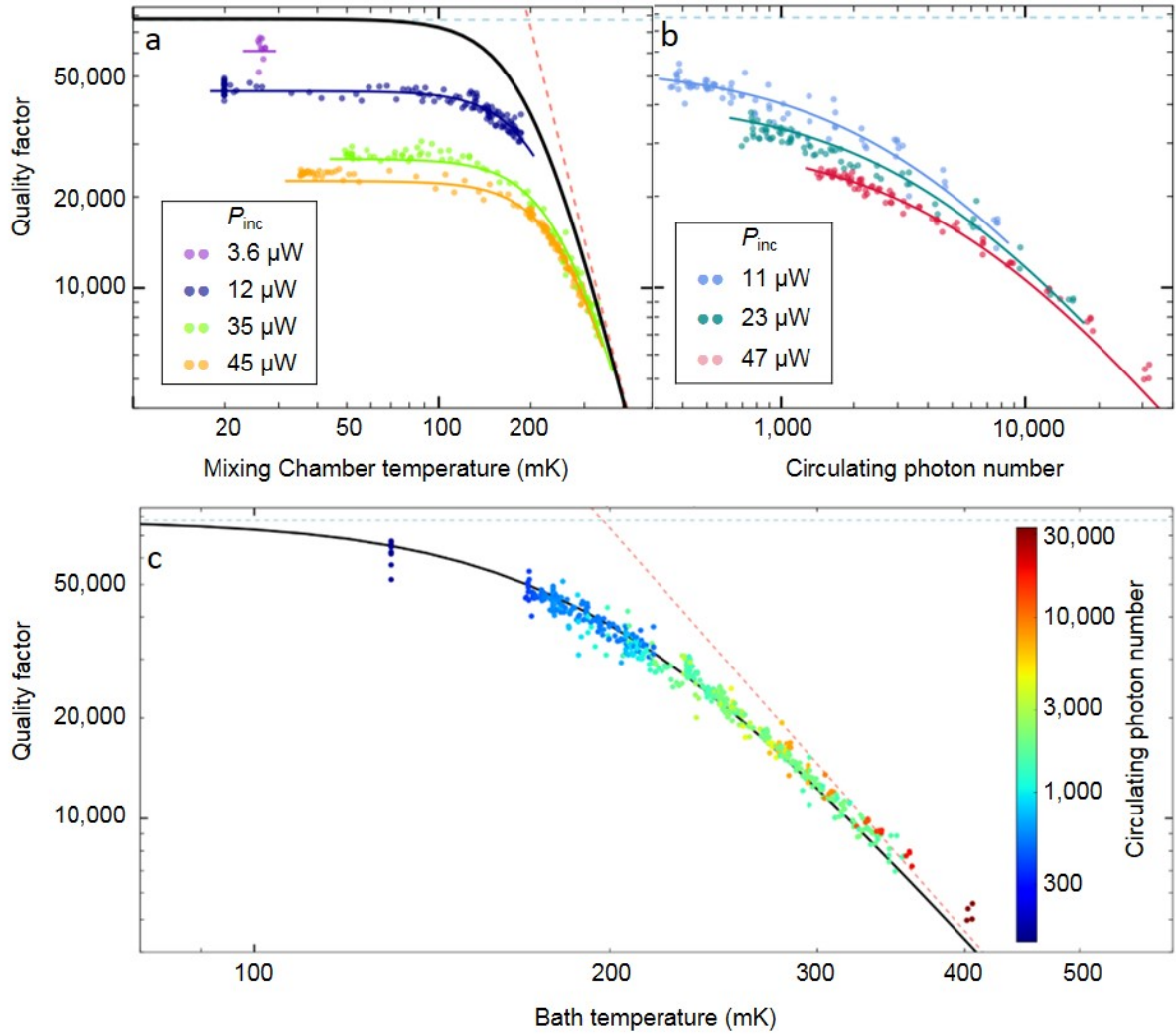

**Figure 3**

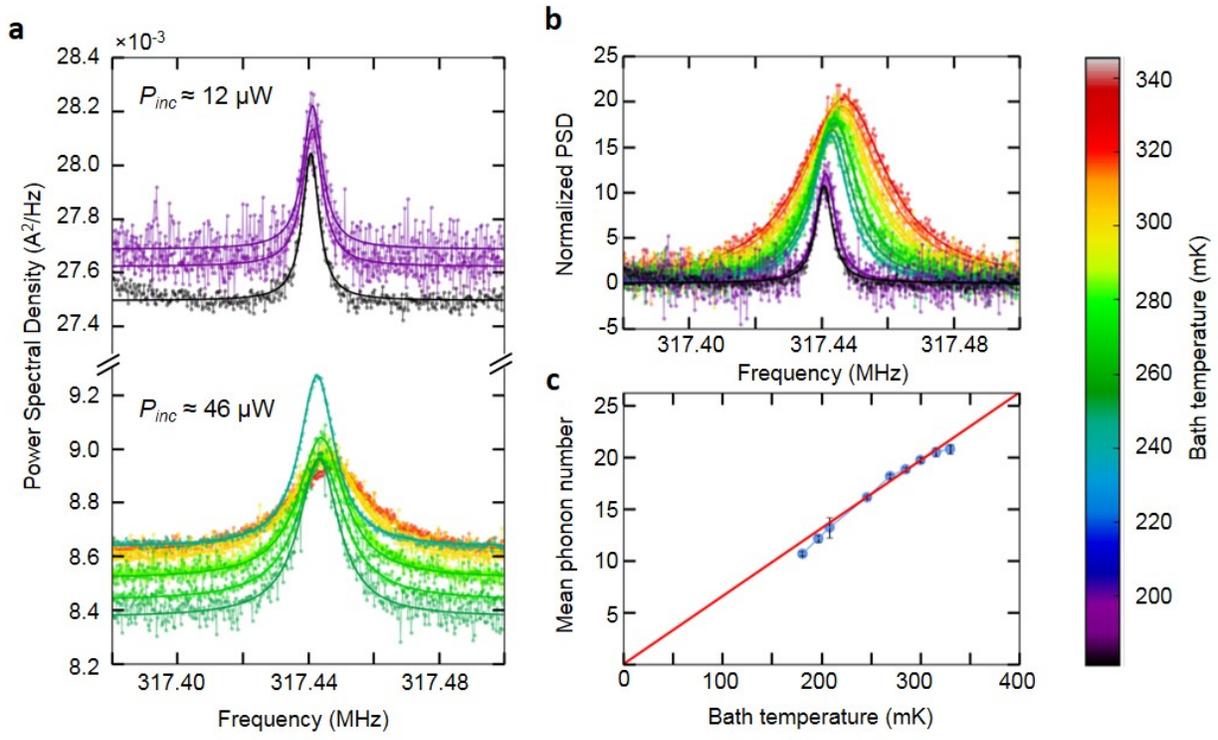

**Figure 4**

# Superfluid Brillouin Optomechanics

A.D. Kashkanova,[1] A.B. Shkarin,[1] C. D. Brown,[1] N. E. Flowers-Jacobs,[1] L. Childress,[1,3] S.W. Hoch,[1] L. Hohmann,[3] K. Ott,[3] J. Reichel,[3] and J. G. E. Harris[1,4]

[1]Department of Physics, Yale University, New Haven, CT, 06511, USA
[2]Department of Physics, McGill University, 3600 Rue University, Montreal, Quebec H3A 2T8, Canada
[3]Laboratoire Kastler Brossel, ENS/UPMC-Paris 6/CNRS, F-75005 Paris, France
[4]Department of Applied Physics, Yale University, New Haven, CT, 06511, USA

## SI A  OMIT/OMIA measurements

### SI A.1  Theory

In this section we describe the theory of Optomechanically Induced Transparency and Optomechanically Induced Amplification (OMIT/OMIA) measurements when a slow photothermal process exists in addition to the usual optomechanical coupling. Using the same notation as the main paper, the standard optomechanics Hamiltonian is

$$\hat{\mathcal{H}} = \hbar\omega_\alpha \hat{a}_\alpha^\dagger \hat{a}_\alpha + \hbar\omega_\beta \hat{b}_\beta^\dagger \hat{b}_\beta + \hbar g_0^{\alpha,\beta}(\hat{b}_\beta + \hat{b}_\beta^\dagger)\hat{a}_\alpha^\dagger \hat{a}_\alpha + \hbar\sqrt{\kappa_{\alpha,\text{in}}}(s_{\text{in}}^* \hat{a}_\alpha + s_{\text{in}} \hat{a}_\alpha^\dagger) \qquad (1)$$

where $\kappa_{\alpha,in}$ is the input coupling and $s_{in}$ is the amplitude of the incoming laser beam

$$s_{\text{in}} = (\bar{s}_{\text{in}} + \delta s_{\text{in}}(t))e^{-i\omega_\text{L} t} \qquad (2)$$

Here $\bar{s}_{\text{in}}$ is a strong "control" beam and $\delta s_{\text{in}}(t)$ is a weak "probe" beam.

The Heisenberg equations of motion are

$$\dot{\hat{a}}_\alpha = -i(\omega_\alpha \hat{a}_\alpha + g_0^{\alpha,\beta}(\hat{b}_\beta + \hat{b}_\beta^\dagger)\hat{a}_\alpha + \sqrt{\kappa_{\alpha,\text{in}}}s_{\text{in}}) - \frac{\kappa_\alpha}{2}\hat{a}_\alpha \qquad (3)$$

and:

$$\dot{\hat{b}}_\beta = -i(\omega_\beta \hat{b}_\beta + g_0^{\alpha,\beta}\hat{a}_\alpha^\dagger \hat{a}_\alpha) - \frac{\gamma_\beta}{2}\hat{b}_\beta - ig_\text{T}\delta T \qquad (4)$$

with the damping rates $\kappa_\alpha$ and $\gamma_\beta$ for the cavity field and the acoustic mode correspondingly. Since OMIT/OMIA is a coherent measurement, the thermal noise of the acoustic mode (which is incoherent) can be neglected. The last term in equation 4 describes the driving of the acoustic mode by changes in the helium temperature $\delta T$. The acoustic mode is coupled to the temperature fluctuations with the coupling rate $g_T$ which has units of Hz/K. Furthermore, we assume that $\delta T$ undergoes simple relaxation towards an equilibrium value set by the intracavity photon number.

$$\dot{\delta T} = g_{\text{ta}}\hat{a}_\alpha^\dagger \hat{a}_\alpha - \kappa_{\text{th}}\delta T \qquad (5)$$

In this equation $g_{\text{ta}}$ is the single photon heating rate in units of K/s, and $\kappa_{th}$ is the relaxation rate of the temperature of the helium inside the cavity.

The two beams produce intensity beating with frequency $\Omega_D \approx \omega_\beta$ and with amplitude $A$, so

$$\hat{a}_\alpha^\dagger \hat{a}_\alpha = \bar{n}_\alpha + (Ae^{-i\Omega_D t} + \text{c.c.}) \tag{6}$$

Here $\bar{n}_\alpha$ is the average circulating photon number. We combine equation 6 with equation 5 and solve for $\delta T$, ignoring the constant temperature shift (due to $\bar{n}_\alpha$). The solution is

$$\delta T = \frac{g_{\text{ta}}}{\kappa_{\text{th}}} \left( \frac{Ae^{-i\Omega_D t}}{1 - i\Omega_D/\kappa_{\text{th}}} + \text{c.c.} \right) \tag{7}$$

Plugging this back into equation 4 and ignoring the counter-rotating term proportional to $e^{+i\Omega_D t}$ gives

$$\dot{\hat{b}}_\beta = -i\left( \omega_\beta \hat{b}_\beta + g_0^{\alpha,\beta}\bar{n}_\alpha + \left( g_0^{\alpha,\beta} + g_T \frac{g_{\text{ta}}}{\kappa_{\text{th}}} \frac{1}{1 - i\Omega_D/\kappa_{\text{th}}} \right) Ae^{-i\Omega_D t} \right) - \frac{\gamma_\beta}{2}\hat{b}_\beta \tag{8}$$

We now define

$$g_{0,\text{pt}}^{\alpha,\beta} = g_T \frac{g_{\text{ta}}}{\kappa_{\text{th}}} \frac{1}{i + \Omega_D/\kappa_{\text{th}}} \approx g_T \frac{g_{\text{ta}}}{\Omega_D} \approx g_T \frac{g_{\text{ta}}}{\omega_\beta}, \tag{9}$$

where we've assumed that the thermal relaxation rate $\kappa_{\text{th}}$ is much smaller than the drive frequency $\Omega_D$. The total coupling to the betanote part of the intracavity power is $g_{\text{tot}} = g_0^{\alpha,\beta}(1 + ig_{0,\text{pt}}^{\alpha,\beta}/g_0^{\alpha,\beta}) = g_0^{\alpha,\beta}G$, where $G$ is defined as

$$G = 1 + i\frac{g_{0,\text{pt}}^{\alpha,\beta}}{g_0^{\alpha,\beta}} \tag{10}$$

With this modification, equation 4 becomes

$$\dot{\hat{b}}_\beta = -i(\omega_\beta \hat{b}_\beta + g_0^{\alpha,\beta}\bar{n}_\alpha + g_0^{\alpha,\beta}GAe^{-i\Omega_D t}) - \frac{\gamma_\beta}{2}\hat{b}_\beta \tag{11}$$

Note that the equation of motion for $\hat{a}_\alpha$ (equation 3) is not influenced by the inclusion of this photothermal process, because the action of the acoustic mode on the intracavity light is only via changes in the He density, which are fully parameterized by $\hat{b}_\beta$.

In the frame rotating at $\omega_L$, defining $\Delta = \omega_L - \omega_\alpha$, and redefining $\hat{a}_\alpha$ and $s_{\text{in}}$ to be rotating at $\omega_L$ as well, we have

$$\dot{\hat{a}}_\alpha = i\Delta\hat{a}_\alpha - ig_0^{\alpha,\beta}(\hat{b}_\beta + \hat{b}_\beta^\dagger)\hat{a}_\alpha - i\sqrt{\kappa_{\alpha,\text{in}}}s_{\text{in}} - \frac{\kappa_\alpha}{2}\hat{a}_\alpha \tag{12}$$

The steady state equations are

$$\bar{a}_\alpha = \frac{-i\sqrt{\kappa_{\alpha,\text{in}}}\bar{s}_{\text{in}}}{-i\bar{\Delta} + \frac{\kappa_\alpha}{2}} \tag{13}$$

and

$$\bar{b}_\beta = \frac{-ig_0^{\alpha,\beta}|\bar{a}_\alpha|^2}{i\omega_\beta + \frac{\gamma_\beta}{2}} \tag{14}$$

where an effective detuning is defined: $\bar{\Delta} = \Delta - g_0^{\alpha,\beta}(\bar{b}_\beta + \bar{b}_\beta^*)$.

Now we linearize the equations around the mean optical amplitude $\bar{a}$, mean acoustic amplitude $\bar{b}$ and average laser drive $\bar{s}_{in}$, by assuming $\hat{a}_\alpha(t) = \bar{a}_\alpha + \delta\hat{a}_\alpha(t)$, $\hat{b}_\beta(t) = \bar{b}_\beta + \delta\hat{b}_\beta(t)$ and $s_{in} = \bar{s}_{in} + \delta s_{in}(t)$. Plugging this assumption into equations 12 and 4 and making use of equations 13 and 14 we have

$$\delta\dot{\hat{a}}_\alpha(t) = i\bar{\Delta}\delta\hat{a}_\alpha(t) - ig_0^{\alpha,\beta}(\delta\hat{b}_\beta(t) + \delta\hat{b}_\beta^\dagger(t))\bar{a}_\alpha - i\sqrt{\kappa_{\alpha,\text{in}}}\delta s_{\text{in}}(t) - \frac{\kappa_\alpha}{2}\delta\hat{a}_\alpha(t) \quad (15)$$

$$\delta\dot{\hat{b}}_\beta(t) = -i\omega_\beta \delta\hat{b}_\beta(t) - ig_0^{\alpha,\beta}G(\bar{a}^*\delta\hat{a}_\alpha(t) + \bar{a}\delta\hat{a}_\alpha^\dagger(t)) - \frac{\gamma_\beta}{2}\delta\hat{b}_\beta \quad (16)$$

In the equation 16 we recognized that $\bar{a}^*\delta\hat{a}_\alpha(t) + \bar{a}\delta\hat{a}_\alpha^\dagger(t)$ is the term corresponding to the intracavity power beatnote that we've denoted as $(Ae^{-i\Omega_D t} + \text{c.c.})$ earlier, so its optomechanical coupling should have the photothermal correction factor $G$. The Fourier transform of equation 15 is

$$-i\omega\delta\hat{a}_\alpha[\omega] = i\bar{\Delta}\delta\hat{a}_\alpha[\omega] - ig_0^{\alpha,\beta}(\delta\hat{b}_\beta[\omega] + \delta\hat{b}_\beta^\dagger[\omega])\bar{a}_\alpha - i\sqrt{\kappa_{\alpha,\text{in}}}\delta s_{\text{in}}[\omega] - \frac{\kappa_\alpha}{2}\delta\hat{a}_\alpha[\omega] \quad (17)$$

By defining the cavity susceptibility

$$\chi_{\text{cav}}[\omega] = \frac{1}{-i\omega - i\bar{\Delta} + \frac{\kappa_\alpha}{2}} \quad (18)$$

and multiphoton coupling

$$g = g_0^{\alpha,\beta}\bar{a}_\alpha \quad (19)$$

we can rewrite equation 17 and its complex conjugate as

$$\delta\hat{a}_\alpha[\omega] = -i\chi_{\text{cav}}[\omega]\left(g(\delta\hat{b}_\beta[\omega] + \delta\hat{b}_\beta^\dagger[\omega]) + \sqrt{\kappa_{\alpha,\text{in}}}\delta s_{\text{in}}[\omega]\right) \quad (20)$$

$$\delta\hat{a}_\alpha^\dagger[\omega] = i\chi_{\text{cav}}^*[-\omega]\left(g^*(\delta\hat{b}_\beta[\omega] + \delta\hat{b}_\beta^\dagger[\omega]) + \sqrt{\kappa_{\alpha,\text{in}}}\delta s_{\text{in}}^*[\omega]\right) \quad (21)$$

Now we take the Fourier transform of equation 16

$$-i\omega\delta\hat{b}_\beta[\omega] = -i\omega_\beta\delta\hat{b}_\beta[\omega] - iG(g^*\delta\hat{a}_\alpha[\omega] + g\delta\hat{a}_\alpha^\dagger[\omega]) - \frac{\gamma_\beta}{2}\delta\hat{b}_\beta[\omega] \quad (22)$$

Combining equations 20 and 21

$$\begin{aligned}(g^*\delta\hat{a}_\alpha[\omega] + g\delta\hat{a}_\alpha^\dagger[\omega]) =& i|g|^2(\delta\hat{b}_\beta[\omega] + \delta\hat{b}_\beta^\dagger[\omega])(\chi_{\text{cav}}^*[-\omega] - \chi_{\text{cav}}[\omega]) \\ &+ i\sqrt{\kappa_{\alpha,\text{in}}}(\chi_{\text{cav}}^*[-\omega]\delta s_{\text{in}}^*[\omega]g - \chi_{\text{cav}}[\omega]\delta s_{\text{in}}[\omega]g^*)\end{aligned} \quad (23)$$

Since $\gamma_\beta \ll \omega_\beta$, we can neglect the counter-rotating term $\delta\hat{b}_\beta^\dagger[\omega]$, which is peaked around $-\omega_\beta$, so

$$\delta\hat{b}_\beta[\omega] = \frac{G\sqrt{\kappa_{\alpha,\text{in}}}(\chi_{\text{cav}}^*[-\omega]\delta s_{\text{in}}^*[\omega]g - \chi_{\text{cav}}[\omega]\delta s_{\text{in}}[\omega]g^*)}{-i(\omega - \omega_\beta) + \frac{\gamma_\beta}{2} + G|g|^2(\chi_{\text{cav}}[\omega] - \chi_{\text{cav}}^*[-\omega])} \quad (24)$$

Defining

$$i\Sigma[\omega] = G|g|^2(\chi_{\text{cav}}[\omega] - \chi_{\text{cav}}^*[-\omega]) \quad (25)$$

results in
$$\delta \hat{b}_\beta[\omega] = \frac{G\sqrt{\kappa_{\alpha,\text{in}}}(\chi^*_{\text{cav}}[-\omega]\delta s^*_{\text{in}}[\omega]g - \chi_{\text{cav}}[\omega]\delta s_{\text{in}}[\omega]g^*)}{-i(\omega-\omega_\beta) + \frac{\gamma_\beta}{2} + i\Sigma[\omega]} \tag{26}$$

In this notation the optical spring $\Delta\omega_{\beta(\text{opt})}$ and optical damping $\gamma_{\beta(\text{opt})}$ are correspondingly

$$\Delta\omega_{\beta(\text{opt})} = \text{Re}[\Sigma[\omega]] \tag{27}$$
$$\gamma_{\beta(\text{opt})} = -2\text{Im}[\Sigma[\omega]] \tag{28}$$

Now we calculate the amplitude and phase of the OMIA signal. In order to observe OMIT/OMIA it is necessary to have two beams detuned by a frequency nearly equal to that of the acoustic oscillator. As shown by equation 2, there are two beams in the system: a strong control beam and a weak probe beam. The probe beam is detuned from the control beam by $-\Omega$, where $\Omega > 0$. In the rotating frame, the expression for $\delta s_{\text{in}}(t)$ is

$$\delta s_{\text{in}}(t) = s_\text{p} e^{i\Omega t} \tag{29}$$

Taking the Fourier transform and assuming $s_p$ to be real

$$\delta s_{\text{in}}[\omega] = \sqrt{2\pi}s_\text{p}\delta(\omega+\Omega) \tag{30}$$

Putting this back into equation 24

$$\delta \hat{b}_\beta[\omega] = \frac{\sqrt{2\pi}s_\text{p} G\sqrt{\kappa_{\alpha,\text{in}}}(g\chi^*_{\text{cav}}[-\omega]\delta(\omega-\Omega) - g^*\chi_{\text{cav}}[\omega]\delta(\omega+\Omega))}{-i(\omega-\omega_\beta) + \frac{\gamma_\beta}{2} + i\Sigma[\omega]} \tag{31}$$

Going back into the time domain

$$\delta \hat{b}_\beta(t) = b_+[\Omega]s_\text{p} e^{-i\Omega t} + b_-[\Omega]s_\text{p} e^{i\Omega t} \tag{32}$$

where

$$b_+[\Omega] = \frac{G\sqrt{\kappa_{\alpha,\text{in}}}\chi^*_{\text{cav}}[-\Omega]g}{-i(\Omega-\omega_\beta) + \frac{\gamma_\beta}{2} + i\Sigma[\Omega]} \tag{33}$$

and

$$b_-[\Omega] = \frac{G\sqrt{\kappa_{\alpha,\text{in}}}(-\chi_{\text{cav}}[-\Omega]g^*)}{-i(-\Omega-\omega_\beta) + \frac{\gamma_\beta}{2} + i\Sigma[-\Omega]} \tag{34}$$

The expression for $b_+$ gives the complex amplitude of the acoustic oscillator. Oscillating at $-\Omega$, $b_-$ is far off resonance, so it will be small.

The optical amplitude is given by equation 20 as

$$\delta \hat{a}_\alpha[\omega] = -i\chi_{\text{cav}}[\omega]\left(g(\delta\hat{b}_\beta[\omega] + \delta\hat{b}^\dagger_\beta[\omega]) + \sqrt{\kappa_{\alpha,\text{in}}}\delta s_{\text{in}}[\omega]\right) \tag{35}$$

Recasting the acoustic mode amplitude in the time domain and neglecting $b_-$ yields

$$g(\delta\hat{b}_\beta(t) + \delta\hat{b}^\dagger_\beta(t)) + \sqrt{\kappa_{\alpha,\text{in}}}\delta s_{\text{in}}(t) = \left(gb_+[\Omega]e^{-i\Omega t} + \left(gb^*_+[\Omega] + \sqrt{\kappa_{\alpha,\text{in}}}\right)e^{i\Omega t}\right)s_\text{p} \tag{36}$$

We express the cavity mode amplitude as

$$\delta\hat{a}_\alpha(t) = a_+[\Omega]e^{-i\Omega t} + a_-[\Omega]e^{i\Omega t} \tag{37}$$

where
$$a_+[\Omega] = -i\chi_{\text{cav}}[\Omega]gb_+[\Omega]s_{\text{p}} \qquad (38)$$

and
$$a_-[\Omega] = -i\chi_{\text{cav}}[-\Omega]\left(gb_+^*[\Omega] + \sqrt{\kappa_{\alpha,\text{in}}}\right)s_{\text{p}} \qquad (39)$$

In the measurement scheme employed in the experiment (having a probe beam only at $-\Omega$) only $a_-$ can be measured. In figure 2 of the main paper, we plot $a_-$ normalized with respect to the background. This normalized signal $a'_-$ is given by:

$$a'_-[\Omega] = \frac{a_-[\Omega]}{a_-[\infty]} = \frac{gb_+^*[\Omega]}{\sqrt{\kappa_{\alpha,\text{in}}}} + 1 \qquad (40)$$

The functions $a[\Omega]$ and $\psi[\Omega]$ are the magnitude and phase of $a'_-[\Omega]$. The values $A$ and $\Psi$, described in the main paper are found as follows

$$A = \text{abs}\left[\frac{gb_+^*[\omega_\beta]}{\sqrt{\kappa_{\alpha,\text{in}}}}\right] \qquad (41)$$

$$\Psi = \arg\left[\frac{gb_+^*[\omega_\beta]}{\sqrt{\kappa_{\alpha,\text{in}}}}\right] \qquad (42)$$

Where $b_+^*[-\omega_\beta]$ is the complex amplitude of the acoustic oscillator, when the the probe beam is detuned by $-\omega_\beta$ from the control beam.

## SI A.2  Full treatment of phase modulation.

In order to use the treatment above to describe the measurements discussed in the main text, we need to take into account some additional features of the measurement setup.

First, as mentioned in the main text, the probe and the control beams are generated in a phase modulator. Specifically, two microwave tones drive the phase modulator: a stronger one at frequency $\omega_{\text{control}}$, and a weaker one at $\omega_{\text{probe}} = \omega_{\text{control}} + \Omega$ (in the actual experiment we also send a third, even weaker, tone that is used for locking the laser to the cavity; its power is at least 3 times lower than the probe tone and it is at a different frequency, so it has a negligible effect on the measurement result). The strong and the weak microwave tones generate the control and the probe optical tones respectively, while the optical carrier acts as a local oscillator in the heterodyne measurement. However, phase modulation generates sidebands symmetrically about the carrier, resulting in negative-order sidebands on the other side of the carrier. These are far enough detuned from the cavity that they don't noticeably affect the mechanical motion; nevertheless, their beatnotes with the carrier are phase-coherent with the beatnotes produced by the control and probe beams, so they will partially cancel the expected heterodyne signal, thus influencing the measurement result.

Moreover, in some of the measurements the microwave tones are strong enough that we need to take into account higher-order optical sidebands (e.g., at frequencies $2\omega_{\text{control}}$ or $\omega_{\text{control}} + \omega_{\text{probe}}$). These can contribute to the beatnote in the photocurrent, and can also be close enough to the cavity resonance to influence mechanical motion.

Finally, the local oscillator is not infinitely far detuned from the cavity (in our case, it is detuned by $\sim 15\kappa$). Hence, it experiences some cavity-induced phase shift in reflection, which also needs to be taken into account.

In order to consistently account for all the factors listed above, we start with the description of the phase modulator. We denote the incident optical tone by $a_0 e^{-i\omega_0 t}$, and a set of microwave tones by $\phi_n \cos(\omega_n t)$, where the $\phi_n$ are the microwave amplitudes normalized by the (possibly frequency-dependent) $V_\pi$ of the phase modulator, and the $\omega_n$ are the corresponding microwave frequencies. The output of the modulator is then expressed as

$$a_\phi = a_0 e^{-i\omega_0 t} e^{i \sum_n \phi_n \cos(\omega_n t)} = a_0 e^{-i\omega_0 t} \prod_n e^{i\phi_n \cos(\omega_n t)} \tag{43}$$

Next, we use the Jacobi-Anger expansion for the exponents

$$e^{i\phi_n \cos(\omega_n t)} = \sum_{k=-\infty}^{+\infty} (-i)^k J_k(\phi_n) e^{-ik\omega_n t}, \tag{44}$$

where $J_k(z)$ is $k^{\text{th}}$ Bessel function of the first kind. With this, the output laser becomes

$$a_\phi = a_0 e^{-i\omega_0 t} \prod_n \left[ \sum_{k=-\infty}^{+\infty} (-i)^k J_k(\phi_n) e^{-ik\omega_n t} \right] \equiv e^{-i\omega_0 t} \sum_l a_l e^{-i\omega_l t}, \tag{45}$$

where $\omega_l$ are all possible intermodulation frequencies resulting from the $\omega_n$. In practice, to keep the computation time short we limit our calculations to a finite intermodulation order (as discussed below).

Next, we consider all of these beams (including the carrier) landing on the cavity. To recap equations 12 and 11, the equations of motion can be written as

$$\dot{a}_\alpha = -(\kappa_\alpha/2 + i\omega_\alpha)a_\alpha - ig_0^{\alpha,\beta} a_\alpha (b_\beta + b_\beta^*) - i\sqrt{\kappa_{\alpha,\text{in}}} s_\text{in} \tag{46}$$

$$\dot{b}_\beta = -(\gamma_\beta/2 + i\omega_\beta)b_\beta - ig_0^{\alpha,\beta} G a_\alpha^* a_\alpha \tag{47}$$

In the frame rotating at the carrier frequency $\omega_0$ the incident optical field becomes $s_\text{in} = \sum_l a_l e^{-i\omega_l t}$, and the optical equation of motion can be written as

$$\dot{a}_\alpha = -(\kappa_\alpha/2 - i\Delta)a_\alpha - ig_0^{\alpha,\beta} a_\alpha (b_\beta + b_\beta^*) - i\sqrt{\kappa_{\alpha,\text{in}}} \sum_l a_l e^{-i\omega_l t}, \tag{48}$$

with $\Delta = \omega_0 - \omega_\alpha$ being the carrier (i.e., local oscillator) detuning.

As usual, next we linearize this equation. To zeroth order in $g_0^{\alpha,\beta}$, the field amplitude is

$$a_0 = \sum_l a_{l,0} e^{-i\omega_l t} \tag{49}$$

$$a_{l,0} = -i\sqrt{\kappa_{\alpha,\text{in}}} a_l \chi_\text{cav}[\omega_l], \tag{50}$$

where $\chi_\text{cav}[\omega] = (\kappa_\alpha/2 - i(\omega + \Delta))^{-1}$. This amplitude results in a force on the mechanical oscillator given by

$$F_0 = -ig_0^{\alpha,\beta} G a_0^* a_0 = -ig_0^{\alpha,\beta} G \sum_l \sum_k a_{l,0} a_{k,0}^* e^{-i(\omega_l - \omega_k)t} \tag{51}$$

The first order in $g_0^{\alpha,\beta}$ terms due to the mechanical motion are (in the adiabatic regime, where $\gamma_\beta \ll \kappa_\alpha$)

$$a_1 = -ig_0^{\alpha,\beta} \sum_l a_{l,0} e^{-i\omega_l t} \left( b_\beta \chi_{\text{cav}}[\omega_l + \omega_\beta] + b_\beta^* \chi_{\text{cav}}[\omega_l - \omega_\beta] \right) \tag{52}$$

Their contribution to the mechanical force is

$$\begin{aligned} F_1 &= -ig_0^{\alpha,\beta} G (a_0^* a_1 + a_0 a_1^*) \\ &= (g_0^{\alpha,\beta})^2 G \sum_l \sum_k a_{l,0} a_{k,0}^* e^{-i(\omega_l - \omega_k)t} \left[ b_\beta \left( \chi_{\text{cav}}^*[\omega_k - \omega_\beta] - \chi_{\text{cav}}[\omega_l + \omega_\beta] \right) \right. \\ &\quad \left. + b_\beta^* \left( \chi_{\text{cav}}^*[\omega_k + \omega_\beta] - \chi_{\text{cav}}[\omega_l - \omega_\beta] \right) \right] \end{aligned} \tag{53}$$

We can make a couple of simplifying assumptions. First, we note that $b_\beta$ rotates at $+\omega_\beta$, while $b_\beta^*$ is located around $-\omega_\beta$. This means that they will be coupled only by the terms rotating at $2\omega_\beta$ (i.e., for $|\omega_k - \omega_l| \approx 2\omega_\beta$). In our case, these terms should be very small, since they would only come from the higher (at least, fourth) order in sideband amplitudes. Thus, we can neglect the $b_\beta^*$ term and get

$$\begin{aligned} F_1 &= -i\Sigma b_\beta \tag{54} \\ \Sigma &= i(g_0^{\alpha,\beta})^2 G \sum_l \sum_k a_{l,0} a_{k,0}^* e^{-i(\omega_l - \omega_k)t} \left( \chi_{\text{cav}}^*[\omega_k - \omega_\beta] - \chi_{\text{cav}}[\omega_l + \omega_\beta] \right) \tag{55} \end{aligned}$$

Second, we are predominantly interested in the DC terms in $\Sigma$; everything else will result in terms rotating at frequencies other than $\omega_\beta$ and, again, will end up far away from the acoustic resonance. As a result, the acoustic equation of motion becomes

$$\begin{aligned} \dot{b}_\beta &= -(\gamma_\beta/2 + i\omega_\beta + i\Sigma) b_\beta + F_0 \tag{56} \\ \Sigma &= i(g_0^{\alpha,\beta})^2 G \sum_l |a_{l,0}|^2 \left( \chi_{\text{cav}}^*[\omega_l - \omega_\beta] - \chi_{\text{cav}}[\omega_l + \omega_\beta] \right) \tag{57} \end{aligned}$$

The solution for this equation is

$$\begin{aligned} b_\beta &= -i(g_0^{\alpha,\beta})^2 G \sum_l \sum_k \chi_{\beta,\text{eff}}[\omega_l - \omega_k] a_{l,0} a_{k,0}^* e^{-i(\omega_l - \omega_k)t} \tag{58} \\ \chi_{\beta,\text{eff}}[\omega] &= (\gamma_\beta/2 + i\omega_\beta + i\Sigma - i\omega)^{-1} \tag{59} \end{aligned}$$

Finally, knowing this expression for $b_\beta$ we can obtain the intracavity power using the first-order solution (equation (52)). Some of the terms in $a_1$ will end up being phase-coherent with the initial optical drive (e.g., the motional sideband on the control beam will be phase-coherent with the probe beam, and vice versa). It is these terms that will define the OMIT response.

For the analysis presented in the main body of the paper, we implement this procedure as follows

- First, knowing the microwave drives and phase modulator response, calculate the tones on the output of the phase modulator using equation 45 (for practical reasons, we limit expansion to the third order in the control beam amplitude and to the first order in the probe beam amplitude);

- Next, from equations 49 and 50 determine the zeroth order intracavity field;

- Use this to calculate the optical force (equation 51) and self-energy (equation 57);

- From these, determine the mechanical response (equations 58 and 59);

- Use this calculated response to get the resulting field inside the cavity $a_\alpha = a_0 + a_1$, where $a_1$ is determined by equation 52

- The reflected field is found using the standard input-output relations $a_{\alpha,\text{out}} = a_{\alpha,\text{in}} - i\sqrt{\kappa_{\alpha,\text{in}}}a_\alpha$;

- Finally, to relate this field to the measurable quantities, we calculate the electrical response of the photodiode $I \propto a^*_{\alpha,\text{out}} a_{\alpha,\text{out}}$, which will consist of all possible beatnotes of the reflected optical tones. The amplitude of the electrical tone oscillating at $\omega_{\text{probe}}$ (which mainly comes of the beating of the probe beam with the carrier) is the relevant OMIT signal.

All of these calculation are performed numerically.

## SI B    Optical and acoustic transmission of the cavity mirrors

### SI B.1    General considerations

A cavity made of two lossless mirrors with power transmittances $T_1$ and $T_2$ has finesse [1]

$$\mathcal{F} = \frac{2\pi}{T_1 + T_2} \tag{60}$$

The corresponding quality factor is

$$Q = \frac{f}{\gamma} = \mathcal{F}\frac{f}{FSR} = \frac{\mathcal{F}f(2L)}{v} = \frac{4\pi fL}{v(T_1 + T_2)} \tag{61}$$

Here $f$ and $\gamma$ are the frequency and linewidth of the cavity mode, $L$ is the cavity length and $v$ is the speed of wave propagation (either optical or acoustic). A conventional way of creating mirrors with very low loss and low transmission is by using Distributed Bragg Reflectors (DBRs).

### SI B.2    The optical transmission through a DBR

This brief review follows the treatment in ref.[2]. Figure 1 shows a schematic representation of a DBR, where the amplitude of the incident field is $E^{(i)} = 1$, the amplitudes of reflected and transmitted fields are $E^{(r)}$ and $E^{(t)}$ correspondingly and the amplitude of the fields propagating forward and backward in the $j^{\text{th}}$ layer are $E_j^{(f)}$ and $E_j^{(b)}$ correspondingly. The index of refraction of the material from which the wave is incident is $n_i$ and the index of refraction of the material into which the wave is transmitted is $n_t$. The index of refraction of the $j^{\text{th}}$ layer is $n_j$ and the thickness of the $j^{\text{th}}$ layer is $d_j$. The $x$-axis is pointing to the right.

Assume a plane wave approaches the DBR at normal incidence. The continuity of parallel components of electric and magnetic fields at each interface results in the following boundary conditions

$$E_j(x,t) = E_{j+1}(x,t) \tag{62}$$

$$\frac{dE_j(x,t)}{dx} = \frac{dE_{j+1}(x,t)}{dx} \tag{63}$$

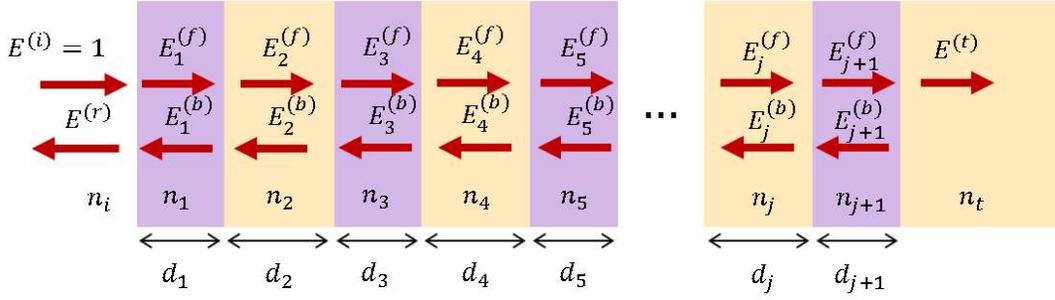

**Figure 1:** DBR showing the electric field in each layer. Color indicates different materials.

At the first interface, the boundary conditions give

$$1 + E^{(\mathrm{r})} = E_1^{(\mathrm{f})} + E_1^{(\mathrm{b})} \tag{64}$$
$$n_i \left(1 - E^{(\mathrm{r})}\right) = n_1 \left(E_1^{(\mathrm{f})} - E_1^{(\mathrm{b})}\right) \tag{65}$$

Applying the boundary conditions to the interface between the $j^{\mathrm{th}}$ and $(j+1)^{\mathrm{th}}$ layers yields the equations

$$E_j^{(\mathrm{f})} e^{ikn_j d_j} + E_j^{(\mathrm{b})} e^{-ikn_j d_j} = E_{j+1}^{(\mathrm{f})} + E_{j+1}^{(\mathrm{b})} \tag{66}$$
$$n_j \left(E_j^{(\mathrm{f})} e^{ikn_j d_j} - E_j^{(\mathrm{b})} e^{-ikn_j d_j}\right) = n_{j+1} \left(E_{j+1}^{(\mathrm{f})} - E_{j+1}^{(\mathrm{b})}\right) \tag{67}$$

Assuming the total number of layers is $p$, applying boundary conditions to the last interface yields

$$E_p^{(\mathrm{f})} e^{ikn_p d_p} + E_p^{(\mathrm{b})} e^{-ikn_p d_p} = E^{(\mathrm{t})} \tag{68}$$
$$n_p \left(E_p^{(\mathrm{f})} e^{ikn_p d_p} - E_p^{(\mathrm{b})} e^{-ikn_p d_p}\right) = n_t E^{(\mathrm{t})} \tag{69}$$

Expressing the equations 64-69 in matrix form

$$\begin{bmatrix} 1 & 1 \\ n_i & -n_i \end{bmatrix} \begin{bmatrix} 1 \\ E^{(\mathrm{r})} \end{bmatrix} = \begin{bmatrix} 1 & 1 \\ n_1 & -n_1 \end{bmatrix} \begin{bmatrix} E_1^{(\mathrm{f})} \\ E_1^{(\mathrm{b})} \end{bmatrix} \tag{70}$$

$$\begin{bmatrix} e^{ikn_j d_j} & e^{-ikn_j d_j} \\ n_j e^{ikn_j d_j} & -n_j e^{-ikn_j d_j} \end{bmatrix} \begin{bmatrix} E_j^{(\mathrm{f})} \\ E_j^{(\mathrm{b})} \end{bmatrix} = \begin{bmatrix} 1 & 1 \\ n_{j+1} & -n_{j+1} \end{bmatrix} \begin{bmatrix} E_{j+1}^{(\mathrm{f})} \\ E_{j+1}^{(\mathrm{b})} \end{bmatrix} \tag{71}$$

$$\begin{bmatrix} e^{ikn_p d_p} & e^{-ikn_p d_p} \\ n_p e^{ikn_p d_p} & -n_p e^{-ikn_p d_p} \end{bmatrix} \begin{bmatrix} E_p^{(\mathrm{f})} \\ E_p^{(\mathrm{b})} \end{bmatrix} = \begin{bmatrix} 1 & 0 \\ n_t & 0 \end{bmatrix} \begin{bmatrix} E^{(\mathrm{t})} \\ 0 \end{bmatrix} \tag{72}$$

Upon rearranging and combining equations 70-72 we arrive at

$$\begin{bmatrix} 1 \\ E^{(\mathrm{r})} \end{bmatrix} = \begin{bmatrix} 1 & 1 \\ n_i & -n_i \end{bmatrix}^{-1} \prod_{j=1}^{p} M_j \begin{bmatrix} 1 & 0 \\ n_t & 0 \end{bmatrix} \begin{bmatrix} E^{(\mathrm{t})} \\ 0 \end{bmatrix} \tag{73}$$

where $M_j$ is defined as

$$M_j = \begin{bmatrix} 1 & 1 \\ n_j & -n_j \end{bmatrix} \begin{bmatrix} e^{ikn_j d_j} & e^{-ikn_j d_j} \\ n_j e^{ikn_j d_j} & -n_j e^{-ikn_j d_j} \end{bmatrix}^{-1} \tag{74}$$

We define:

$$A = \begin{bmatrix} a_{11} & a_{12} \\ a_{21} & a_{22} \end{bmatrix} = \begin{bmatrix} 1 & 1 \\ n_i & -n_i \end{bmatrix}^{-1} \prod_{j=1}^{p} M_j \begin{bmatrix} 1 & 0 \\ n_t & 0 \end{bmatrix} \tag{75}$$

and now we have

$$\begin{bmatrix} 1 \\ E^{(\mathrm{r})} \end{bmatrix} = A \begin{bmatrix} E^{(\mathrm{t})} \\ 0 \end{bmatrix} \tag{76}$$

from which we find

$$r_{\mathrm{opt}} = E^{(\mathrm{r})} = a_{21}/a_{11} \tag{77}$$

$$T_{\mathrm{opt}} = 1 - r_{\mathrm{opt}}^2 \tag{78}$$

To conclude, finding the transmittivity of the DBR involves calculating the matrix $A$, which is straightforward, provided we know the indexes of refraction $n_\mathrm{i}$ and $n_\mathrm{t}$, and the structure of the DBR, which is fully described by $n$ and $d$ of the alternating DBR layers.

### SI B.3   The acoustic transmission through a DBR

Finding the DBR transmittivity for the sound wave is done in a similar manner [3, 4]. Assume a sound wave with frequency $f$ and wavelength $\lambda$ is incident on a DBR. Equating the displacements ($s(x,t)$) and pressures ($P(x,t) = -K ds/dx$) on both sides of the boundary ($K$ is the bulk modulus) gives

$$s_j(d,t) = s_{j+1}(d,t) \tag{79}$$

$$K_j \frac{ds_j}{dx}(d,t) = K_{j+1} \frac{ds_{j+1}}{dx}(d,t) \tag{80}$$

The fact that acoustic impedance is related to bulk modulus via $Z = K/v = v\rho$, where $\rho$ is the density of the material, and that the frequency of the mode must be constant throughout the DBR, results in three sets of equations that provide information about the amplitude of the acoustic mode at different interfaces. At the first interface

$$1 + s^{(\mathrm{r})} = s_1^{(\mathrm{f})} + s_1^{(\mathrm{b})} \tag{81}$$

$$Z_i(1 - s^{(\mathrm{r})}) = Z_1 \left( s_1^{(\mathrm{f})} - s_1^{(\mathrm{b})} \right) \tag{82}$$

At the interface between the $j^{\mathrm{th}}$ and $(j+1)^{\mathrm{th}}$ layer:

$$s_j^{(\mathrm{f})} e^{ikn_j^{\mathrm{ac}} d_j} + s_j^{(\mathrm{b})} e^{-ikn_j^{\mathrm{ac}} d_j} = s_{j+1}^{(\mathrm{f})} + s_{j+1}^{(\mathrm{b})} \tag{83}$$

$$Z_j \left( s_j^{(\mathrm{f})} e^{ikn_j^{\mathrm{ac}} d_j} - s_j^{(\mathrm{b})} e^{-ikn_j^{\mathrm{ac}} d_j} \right) = Z_{j+1} \left( s_{j+1}^{(\mathrm{f})} - s_{j+1}^{(\mathrm{b})} \right) \tag{84}$$

Here we defined $k = 2\pi/\lambda$ and $n_j^{\text{ac}} = v_i/v_j$, where $v_i$ and $v_j$ are correspondingly the speed of the sound wave in the medium on the incident side and the speed of sound in the $j^{\text{th}}$ layer. At the last interface, assuming a total number of layers $p$

$$s_p^{(\text{f})} e^{ikn_p^{\text{ac}}d_p} + s_p^{(\text{b})} e^{-ikn_p^{\text{ac}}d_p} = s^{(\text{t})} \tag{85}$$

$$Z_p \left( s_p^{(\text{f})} e^{ikn_p^{\text{ac}}d_p} - s_p^{(\text{b})} e^{-ikn_p^{\text{ac}}d_p} \right) = Z_t s^{(\text{t})} \tag{86}$$

From equations 81-86, using the same methods as in section SI B.2 , a matrix $B$ is obtained:

$$B = \begin{bmatrix} b_{11} & b_{12} \\ b_{21} & b_{22} \end{bmatrix} = \begin{bmatrix} 1 & 1 \\ Z_i & -Z_i \end{bmatrix}^{-1} \prod_{j=1}^{p} M_j^{\text{ac}} \begin{bmatrix} 1 & 0 \\ Z_t & 0 \end{bmatrix} \tag{87}$$

where $M_j^{\text{ac}}$ is defined as:

$$M_j^{\text{ac}} = \begin{bmatrix} 1 & 1 \\ Z_j & -Z_j \end{bmatrix} \begin{bmatrix} e^{ikn_j^{\text{ac}}d_j} & e^{-ikn_j^{\text{ac}}d_j} \\ Z_j e^{ikn_j^{\text{ac}}d_j} & -Z_j e^{-ikn_j^{\text{ac}}d_j} \end{bmatrix}^{-1} \tag{88}$$

Now we have:
$$\begin{bmatrix} 1 \\ s^{(\text{r})} \end{bmatrix} = B \begin{bmatrix} s^{(\text{t})} \\ 0 \end{bmatrix} \tag{89}$$

from which we get:
$$r_{\text{ac}} = s^{(\text{r})} = b_{21}/b_{11} \tag{90}$$

$$T_{\text{ac}} = 1 - r_{\text{ac}}^2 \tag{91}$$

In conclusion, to calculate both optical and acoustic reflection/transmission of a DBR the following quantities need to be known:

- $d$ -thickness of each layer
- $n_{\text{opt}}$ - optical index of refraction for each layer
- $\rho$ - density for each layer
- $v$ - sound velocity in each layer

Additionally, it is necessary to know the values of those parameters for the material from which the wave is incident and the material into which the wave is transmitted. For the experiments described in the main paper these are liquid $^4$He and SiO$_2$ respectively.

### SI B.4  Optical and acoustic quality factors for the present DBRs

The DBRs in these experiments are deposited on to the faces of SiO$_2$ fibers. The DBR is comprised of alternating layers of SiO$_2$ and Ta$_2$O$_5$ whose thiknesses are chosen to correspond to one-quarter wavelength of the light (which in vacuum has wavelength $\sim$ 1550 nm). The cavity is formed between two such fibers immersed in superfluid helium; therefore we are interested in the transmission of light and sound from the superfluid helium, through the DBR, and into the fiber.

| Material | $n_{opt}$ | $v$(m/s) | $\rho$ (kg/m$^3$) |
|---|---|---|---|
| $^4$He | 1.0282[5] | 238[5] | 145[5] |
| SiO$_2$ | 1.4746[6] | 5,900 ± 100[7, 8] | 2,200[7] |
| Ta$_2$O$_5$ | 2.0483[6] | 4,500 ± 500[7] | 7,600 ± 600[7] |

**Supplementary Table 1:** Material properties of superfluid $^4$He, SiO$_2$ and Ta$_2$O$_5$

Table 1 shows the relevant parameters for each material.

For the 1550 nm light, a quarter optical wavelength layer of SiO$_2$ is 263 nm and a quarter optical wavelength layer of Ta$_2$O$_5$ is 189 nm. The 1550 nm light has wavelength equal to 1508 nm in liquid $^4$He and therefore couples to the acoustic mode with 754 nm wavelength in liquid $^4$He, whose frequency is 315.7 MHz. For this mode, a quarter acoustic wavelength in SiO$_2$ is 4.7 ± 0.1 $\mu$m and a quarter acoustic wavelength in Ta$_2$O$_5$ is 3.6±0.4 $\mu$m; the large discrepancy between the optical and acoustic wavelengths *in the solid materials* means that the DBRs used in the present device will not provide strong reflectivity for the acoustic waves.

The experiment was performed with the mirror surfaces separated by 84 $\mu$m. The cavity consisted of a low reflectivity (15 layer pairs) input DBR and high reflectivity (18 layer pairs) back DBR. The laser wavelength was 1538 nm; the frequency of the acoustic mode of interest was $\omega_\beta = 2\pi \times 317.3$ MHz.

As mentioned above, the DBRs in the present devices are not designed to be highly reflective for the acoustic waves; nevertheless, the strong contrast in acoustic impedance between $^4$He and the outermost layer of the optical DBR provides fairly large reflectivity. From equation 91 the acoustic transmittivity for the low reflectivity DBR is calculated to be 10,100 ± 400 ppm, while the acoustic transmittivity for the high reflectivity DBR is calculated to be 7,700 ± 1,300 ppm. The uncertainties predominantly come from the material properties of Ta$_2$O$_5$ (see Table 1). These values of the acoustic transmission lead to the acoustic quality factor $Q_{\beta,\text{ext}}$ =79000 ± 5,000 which is slightly higher than the experimentally determined quality factor $Q_{\beta,\text{ext}}$ =70,000 ± 2,000. This difference might be attributed to various factors, such as losses due to mirror misalignment or imprecise knowledge of the DBR structure.

## SI B.5 Improving the acoustic quality factor in future devices

In this section we consider the possibility of improving the acoustic quality factor by adding an acoustic DBR between the substrate and the optical DBR. The proposed structure in shown in figure 2.

We consider a cavity in which optical confinement is again provided by optical DBRs with 15 and 18 pairs, optimized for maximum reflectivity at 1550 nm, and the acoustic confinement is enhanced by additional layers forming an acoustic DBR. Technical aspects of the coating process impose the requirement that the total stack thickness be less than 25 $\mu$m, and that the thickness of each layer of a given material is an integer multiple of the thickness of the thinnest layer of this material[6]. Maximizing $Q_{\beta,\text{ext}}$ for the 315.7 MHz acoustic mode (to which 1550 nm light couples), subject to the above constraints, we find the optimal design for the stack. Table 2 shows stack parameters for the current designs as well as the proposed designs.

We use equations 78 and 91 to calculate the optical and acoustic transmission for the DBRs presented in table 2. The results are shown in table 3. The optical transmission doesn't change appreciably as more acoustic layers are added, as the acoustic layers are outside of the optical mode. At the same time, the acoustic layers decrease the acoustic transmission by more than an order of magnitude. Assuming the mirror separation is 84 $\mu$m (as in the present devices), we calculate the expected acoustic quality factor to be $Q_{\beta,\text{ext}} = 3.3 \pm 2.2 \times 10^6$, where the uncertainty is primarily due to the acoustic parameters of Ta$_2$O$_5$. This calculation assumes a cavity constructed from Device #3 and Device #4 in Tables 2 and 3.

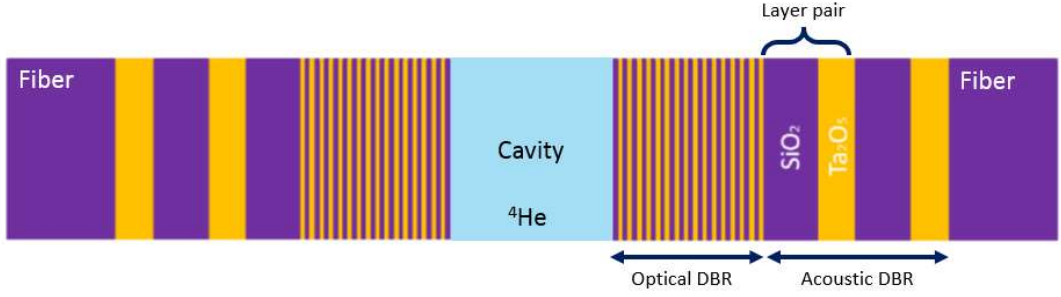

**Figure 2:** Example of a DBR structure that would be reflective for both optical and acoustic waves

| Number | Design |
|---|---|
| 1 (Current) | Substrate×(189 nm Ta$_2$O$_5$×253 nm SiO$_2$)$^{15}$×189 nm Ta$_2$O$_5$ |
| 2 (Current) | Substrate×(189 nm Ta$_2$O$_5$×253 nm SiO$_2$)$^{18}$×189 nm Ta$_2$O$_5$ |
| 3 (Future) | Substrate×3591 nm Ta$_2$O$_5$×4554 nm SiO$_2$×3591 nm Ta$_2$O$_5$×759 nm SiO$_2$× (189 nm Ta$_2$O$_5$×253 nm SiO$_2$)$^{15}$×189 nm Ta$_2$O$_5$ |
| 4 (Future) | Substrate×3780 nm Ta$_2$O$_5$×4554 nm SiO$_2$×3213 nm Ta$_2$O$_5$× (253 nm SiO$_2$×189 nm Ta$_2$O$_5$)$^{18}$ |

**Supplementary Table 2:** The stack designs used in the current device and the future stack designs

| Number | $T_{\text{opt}}$ (ppm) | $T_{\text{ac}}$ (ppm) | $d_{\text{total}}$ ($\mu$m) | $\frac{U^{\text{SiO}_2}}{U^{\text{He}}}$ (ppm) | $\frac{U^{\text{Ta}_2\text{O}_5}}{U^{\text{He}}}$ (ppm) | $Q_{\beta,\text{loss}}$ |
|---|---|---|---|---|---|---|
| 1 | 73.5 | $10,000 \pm 400$ | 6.6 | $6.5 \pm 0.5$ | $17.1 \pm 2.1$ | $4.2 \times 10^7$ |
| 2 | 10.2 | $8,100 \pm 1200$ | 8.3 | $6.4 \pm 0.9$ | $16.8 \pm 1.5$ | $4.3 \times 10^7$ |
| 3 | 73.5 | $350 \pm 200$ | 19.68 | $3.8 \pm 1$ | $9.5 \pm 2$ | $7.5 \times 10^7$ |
| 4 | 10.2 | $420 \pm 240$ | 19.87 | $3.1 \pm 1$ | $7.8 \pm 1.4$ | $9.2 \times 10^7$ |

**Supplementary Table 3:** The calculated parameters for different stacks. The optical transmittivity ($T_{\text{opt}}$), acoustic transmittivity ($T_{\text{ac}}$) and thickness ($d_{\text{total}}$) are discussed in section SI B.5. The ratios of the energy stored in SiO$_2$ ($U^{\text{SiO}_2}/U^{\text{He}}$) and Ta$_2$O$_5$ ($U^{\text{Ta}_2\text{O}_5}/U^{\text{He}}$) to the energy stored in Helium are discussed in SI B.6

## SI B.6  Acoustic loss inside the DBR layers

In addition to being limited by the transmission of the acoustic mode into the fiber, the acoustic Q factor can also be limited by dissipation in the DBR. To estimate the Q factor associated with this loss, the following expression is used

$$\frac{1}{Q_{\beta,\text{loss}}} = \frac{U^{\text{SiO}_2}\phi_{\text{SiO}_2} + U^{\text{Ta}_2\text{O}_5}\phi_{\text{Ta}_2\text{O}_5}}{U^{\text{He}}} \quad (92)$$

Here $U$ is the energy stored in the material; $\phi$ is the acoustic loss angle for the material.

To find the stored energy, we use the treatment above to find the displacement field in each layer. The stored energy can then be expressed as

$$\begin{aligned}
U_j^{\text{stored}} &\propto \int_0^{d_j} \omega_\beta^2 \rho_j |s_j^{(f)} e^{ikn_j^{\text{ac}}x} + s_j^{(b)} e^{-ikn_j^{\text{ac}}x}|^2 dx \quad (93)\\
&= \omega_\beta^2 \rho_j \left[ \left(|s_j^{(f)}|^2 + |s_j^{(b)}|^2\right) d_j - \frac{1}{kn_j^{\text{ac}}} \text{Im}\left(s_j^{(f)*} s_j^{(b)} (e^{-2ikn_j^{\text{ac}} d_j} - 1)\right) \right]
\end{aligned}$$

The calculated ratios of energy stored in SiO$_2$ and Ta$_2$O$_5$ are shown in table 3. Both $\phi_{\text{SiO}_2}$ and $\phi_{\text{Ta}_2\text{O}_5}$ have been measured over a range of temperatures and for frequencies mostly much lower than 300 MHz [9, 10, 11, 12, 13, 14, 15, 16]. All of these measurements show $\phi_{\text{SiO}_2}; \phi_{\text{Ta}_2\text{O}_5} < 10^{-3}$. The values of the quality factor due to absorption in the DBR ($Q_{\beta,\text{loss}}$), assuming $\phi_{\text{SiO}_2} = \phi_{\text{Ta}_2\text{O}_5} = 10^{-3}$, are shown in Table 3. For all designs $Q_{\beta,\text{loss}}$ is much larger than $Q_{\beta,\text{ext}}$. As a result, the acoustic quality factor should not be limited by the acoustic absorption in the DBR.

## SI C  Acoustic loss in the superfluid helium.

For $T < 600$ mK the main intrinsic loss mechanism for density waves (i.e., first sound) in superfluid helium is the three phonon process [17]. It can be described by an amplitude attenuation coefficient $\alpha_{3pp}$

$$\alpha_{3\text{pp}}(\omega_\beta) = \frac{\pi^2 (u+1)^2 k_B^4}{30 \rho_{He} \hbar^3 v_{He}^6} \omega_\beta T^4 \left( \arctan(\omega_\beta \tau) - \arctan\left(\frac{3}{2}\gamma \bar{p}^2 \omega_\beta \tau\right) \right) \quad (94)$$

Here $\omega_\beta$ is the wave frequency, $T$ is the temperature, $\rho_{He} = 145$ kg/m$^3$ and $v_{He} = 238$ m/s are the helium density and sound velocity, $u = 2.84$ is the Grüneisen constant [18], $\tau = \xi T^{-5}$ is the thermal phonon lifetime, where $\xi = 1.11 \times 10^{-7}$ s·K$^5$ [19] and $\bar{p} = 3 k_B T / v_{He}$ is the average thermal phonon momentum. Finally, $\gamma$ is the coefficient for the cubic term in the phonon dispersion, which is expressed as $\gamma = -\frac{1}{6 v_{He}} \frac{d^3 \epsilon}{dp^3}$, where $\epsilon$ and $p$ are phonon energy and momentum respectively. It has been measured to be $\gamma = -8 \times 10^{47}$ kg$^{-2}$m$^{-2}$s$^2$ [18].

The intrinsic quality factor of an acoustic mode can be calculated from the attenuation length as

$$Q_{\beta,\text{int}} = \frac{\omega_\beta}{2 v_{He} \alpha_{3pp}} \quad (95)$$

For the relevant acoustic mode frequency $\omega_\beta = 2\pi \times 313.86$ MHz and temperature $T < 0.5$ K both arctan arguments in equation 94 are $\gg 1$, leading to the simple relationship $Q_{\beta,\text{int}} = \frac{\chi}{T^4}$, where $\chi \approx 118$ K$^4$.

## Heat transport model and thermal response time.

In order to make use of the equations 94 and 95 to analyze the data in the main paper, it is necessary to know the temperature of the helium inside the cavity. To accomplish this, we have developed a model of heat transport in the device. The steady state solution of this model yields the dependence of the device temperature on the dissipated power and the temperature of the mixing chamber, and is used to derive equation 2 in the main text. The dynamical solution provides an expression for the thermal relaxation time of the helium inside the cavity.

## Heat transport equation

First, let us define the geometry of the device. As shown in figure 1a of the main text, a cylindrical volume of helium occupies the space between the faces of two optical fibers which are confined within the bore of a glass ferrule. The helium inside the cavity is thermally linked to a larger volume of helium outside the ferrule via two identical sheaths; since these sheaths have the same length, we can represent them as a single sheath with doubled cross-sectional area. Finally, we assume that the helium outside the ferrule has large heat capacitance and a good thermal link to the mixing chamber, so its temperature doesn't depend on the power dissipated inside the cavity and is the same as the mixing chamber temperature.

We represent the helium inside the cavity as a point heat capacitance $C_0(T)$ located at $x = l$ and experiencing a heat load $\Phi$ (dissipated laser power). This capacitance is connected to a reservoir at $x = 0$ through a one-dimensional channel (the combined sheaths), which has a heat capacitance per unit length $C_l(T)$ and a thermal resistance per unit length $R_l(T)$. The reservoir is maintained at a constant temperature $T_{\mathrm{MC}}$. If we denote the temperature dependent specific heat (per unit volume) of helium by $c_V(T)$ and its thermal conductivity in the channel by $\kappa(T)$, we get for the parameters above

$$C_0(T) = V_{\mathrm{cav}} c_V(T) \tag{96}$$

$$C_l(T) = A_{\mathrm{sh}} c_V(T) \tag{97}$$

$$R_l(T) = (A_{\mathrm{sh}} \kappa(T))^{-1}, \tag{98}$$

where $V_{\mathrm{cav}}$ is the volume of the cylindrical cavity and $A_{\mathrm{sh}}$ is the combined cross-sectional area of the sheaths.

The two equations governing the heat transport in the channel are

$$j = -\frac{1}{R_l(T)}\frac{\partial T}{\partial x} \tag{99}$$

$$C_l(T)\frac{\partial T}{\partial t} = -\frac{\partial j}{\partial x} \tag{100}$$

The first equation relates the heat current $j(x)$ and the temperature gradient $\frac{\partial T}{\partial x}$ (positive values of $j$ denote the heat flowing in the positive $x$ direction, i.e., from the reservoir into the cavity), and the second one describes the heating of the helium inside the channel. The boundary condition at the reservoir is simply $T(x = 0) = T_{\mathrm{MC}}$, while for the cavity it is expressed through a heat flow balance

$$\Phi = \left(C_0 \frac{\partial T}{\partial t} - j\right)\bigg|_{x=l} \tag{101}$$

This last relation shows that the power $\Phi$ dissipated in the cavity is partially spent on increasing its temperature and partially transmitted into the channel.

Because the thermal conductivity $\kappa$ and heat capacity $c_V$ are temperature-dependent, the equations above are in principle non-linear. Nevertheless, since they have the same dependence $c(T) = \delta_V T^3$, $\kappa(T) = \epsilon_V T^3$, we can transform the equations into linear ones with an appropriate substitution. For that, we express the material parameters as

$$C_l(T) = A_{\text{sh}} c(T) = \delta_l T^3 \tag{102}$$

$$C_0(T) = V_{\text{cav}} c(T) = \delta_0 T^3 \tag{103}$$

$$R_l(T) = (A_{\text{sh}} \kappa(T))^{-1} = (\epsilon_l T^3)^{-1}, \tag{104}$$

where $\delta_l = A_{\text{sh}} \delta_V$, $\delta_0 = V_{\text{cav}} \delta_V$ and $\epsilon_l = A_{\text{sh}} \epsilon_V$. Substituting these expressions into the equation and boundary conditions, we obtain

$$j(x) = -\epsilon_l T^3 \frac{\partial T}{\partial x} = -\frac{\epsilon_l}{4} \frac{\partial (T^4)}{\partial x} \tag{105}$$

$$-\frac{\partial j}{\partial x} = \delta_l T^3 \frac{\partial T}{\partial t} = \frac{\delta_l}{4} \frac{\partial (T^4)}{\partial t} \tag{106}$$

$$\Phi = \left( \delta_0 T^3 \frac{\partial T}{\partial t} - j \right) \bigg|_{x=l} = \left( \frac{\delta_0}{4} \frac{\partial (T^4)}{\partial t} - j \right) \bigg|_{x=l} \tag{107}$$

Denoting $u = T^4$ and using the first equation to express $j$ leads to

$$\frac{\partial u}{\partial t} = \frac{\epsilon_l}{\delta_l} \frac{\partial^2 u}{\partial x^2} \tag{108}$$

$$u|_{x=0} = u_0 \tag{109}$$

$$\left( \frac{\partial u}{\partial t} + \frac{\epsilon_l}{\delta_0} \frac{\partial u}{\partial x} \right) \bigg|_{x=l} = \frac{4\Phi}{\delta_0} \tag{110}$$

Thus, the heat transport equation is expressed as a one-dimensional diffusion equation with the diffusion coefficient $D = \epsilon_l / \delta_l$.

**Steady state solution**

First, we consider a steady state solution with a constant heat load $\Phi$. The diffusion equation turns into $\frac{\partial^2 u}{\partial x^2} = 0$, which has a general solution $u = a + bx$. The boundary condition at $x = 0$ immediately yields $a = u_0 = T_{\text{MC}}^4$. From the second boundary condition we find $b = \frac{\partial u}{\partial x} = \frac{4\Phi}{\epsilon_l}$, which results in

$$u(x) = u_0 + \frac{4\Phi}{\epsilon_l} x, \tag{111}$$

From this, the temperature of the cavity can be determined as

$$T_{\text{cav}}^4 = u(l) = T_{\text{MC}}^4 + \frac{4\Phi}{\epsilon_l} l \tag{112}$$

This relation is used to derive equation 2 in the main text.

**Transient dynamics**

Next, we consider the dynamics of this system. We consider the system in the steady state derived above and then abruptly turn off the power source at $t = 0$. The cavity temperature should then decay to $T_0$ exponentially with some characteristic time $\tau_0$, which we want to determine.

To find the time evolution we use the eigenfunctions expansion of the solution:

$$u(x,t) = u_0 + \sum_n T_n(t) v_n(x), \qquad (113)$$

where $v_n(x)$ is an eigenfunction of the Laplace operator with the appropriate boundary conditions

$$\frac{\partial^2 v_n}{\partial x^2} = -\lambda_n v_n \qquad (114)$$

$$v_n(0) = 0 \qquad (115)$$

$$\left(\frac{\epsilon_l}{\delta_l}\frac{\partial^2 v_n}{\partial x^2} + \frac{\epsilon_l}{\delta_0}\frac{\partial v_n}{\partial x}\right)\bigg|_{x=l} = 0 \qquad (116)$$

If we denote $\lambda_n = k_n^2$ (choosing the opposite sign $\lambda = -k_n^2$ results in the inability to satisfy both boundary conditions simultaneously, as well as an exponentially diverging time evolution), we get from the first two equations that $v_n(x) = \sin(k_n x)$. The boundary condition at $x = l$ restricts the values of $k_n$

$$-\frac{\epsilon_l}{\delta_l} k_n^2 \sin(k_n l) + \frac{\epsilon_l}{\delta_0} k_n \cos(k_n l) = 0, \qquad (117)$$

which can be rewritten as

$$(k_n l)\tan(k_n l) = r_V \qquad (118)$$

with $r_V = \frac{l\delta_l}{\delta_0} = \frac{lA_{sh}}{V_{cav}} = \frac{V_{sh}}{V_{cav}}$ is the ratio of the sheath and the cavity volumes. The solutions for this equation exhaust all of the values $k_n$.

Now we can substitute the expansion back into equation 108 to obtain the equations for the time-dependent parts $T_n$:

$$\frac{\partial u}{\partial t} = \frac{\epsilon_l}{\delta_l}\frac{\partial^2 u}{\partial x^2} \qquad (119)$$

$$\sum_n v_n \frac{\partial T_n}{\partial t} = \frac{\epsilon_l}{\delta_l}\sum_n T_n \frac{\partial^2 v_n}{\partial x^2} = -\frac{\epsilon_l}{\delta_l}\sum_n k_n^2 T_n v_n \qquad (120)$$

As the eigenfunctions are orthogonal, equation 120 has to be satisfied for each $T_n$ independently

$$\frac{\partial T_n}{\partial t} = -\frac{\epsilon_l k_n^2}{\delta_l} T_n = -\frac{T_n}{\tau_n}, \qquad (121)$$

where $\tau_n = \frac{\delta_l}{\epsilon_l k_n^2}$ is the characteristic decay time. The solution for this equation is

$$T_n(t) = T_n(0) e^{-t/\tau_n}, \qquad (122)$$

We're mostly interested in the longest relaxation time $\tau_0$ corresponding to the smallest value of $k_n$, which we denote as $k_0$

$$\tau_0 = \frac{\delta_l}{\epsilon_l} \frac{1}{k_0^2} \qquad (123)$$

With several percent error, $k_0$ can be approximated by

$$(k_0 l)^{-2} \approx \left(\frac{2}{\pi}\right)^2 + r_V, \qquad (124)$$

so the relaxation time becomes

$$\tau_0 \approx \frac{\epsilon_l}{\delta_l} l^2 \left(\frac{4}{\pi^2} + r_V\right) = \frac{\delta_l T^3 l}{\epsilon_l T^3/l} \left(\frac{4}{\pi^2} + r_V\right) = \frac{C_{sh}}{\kappa_{sh}} \left(\frac{4}{\pi^2} + r_V\right), \qquad (125)$$

where $C_{sh} = \delta_l T^3 l$ is the total heat capacitance of the sheath, and $\kappa_{sh} = \epsilon_l T^3/l$ is the total thermal conductance of the sheath.

The expression above for the thermal relaxation time $\tau_0$ depends only the sheath's heat capacitance, thermal conductance, and the geometric parameter $r_V$. The heat capacitance of the sheath can be known fairly well, since it only depends on its volume and the specific heat of the liquid helium, which for low temperatures is well known [20]. Thermal conductivity, however, is much harder to evaluate *a priori*, since it depends upon the particular geometry of the conducting channel (which determines the mean phonon travel length between collisions with the boundaries) and the scattering properties of its wall. Therefore, measurements of $\tau_0$ can provide an estimate for the thermal conductivity without the need for any assumptions about the specularity of reflections from sheath surface.

**Measurement of the thermal relaxation time**

We measure $\tau_0$ by changing the circulating optical power (which is proportional to the power dissipated inside the cavity) and monitoring the response of the temperature-dependent linewidth of the acoustic mode. The experiment is performed using the OMIT/OMIA technique described in the main text, but with the probe beam frequency $\omega_{\text{probe}}$ being fixed exactly one acoustic frequency away from the control beam: $\omega_{\text{probe}} = \omega_{\text{control}} + \omega_\beta$. This way, the magnitude of the OMIA part of the probe beam reflection is inversely proportional to the linewidth of the acoustic mode, which is a monotonic function of the device temperature. Hence, by observing the OMIA response as a function of time we can access the temperature dynamics. In practice, rather than measuring a step response to a change in the dissipated optical power, we perform a lock-in measurement where we sinusoidally modulate the optical drive and record the complex response of the magnitude of the OMIA signal.

The results are shown in figure 3. The data was fit to a double exponential decay with two time scales $\tau_s$ and $\tau_f$ and an additional time delay $\tau_d$

$$\delta \gamma_{int}[\omega] \propto \frac{1}{1 + i\omega\tau_s} \frac{1}{1 + i\omega\tau_f} e^{i\omega\tau_d} \qquad (126)$$

We attribute the longer of the two decay times $\tau_s \approx 350$ $\mu$s to the thermal response. The shorter time $\tau_f \approx 40$ $\mu$s is only required to account for the data at frequencies above 2 kHz; it might arise from some other faster thermal process in the system (e.g., heating of the dielectric stack, or thermal equilibration of the helium inside the cavity), or from the time-delayed mechanical response itself. Finally, the time delay $\tau_d \approx 30$ $\mu$s can be attributed to the sound propagation delay, as it is comparable to the ballistic phonon travel time in the sheath $l/v_{He} \approx 12$ $\mu$s. In interpreting the slowest time $\tau_s$ as the thermal response time $\tau_0$ we assumed that the thermal response is the slowest time scale in the system. Indeed, the slower time $\tau_s$ we observe is much larger than either optical ($\kappa_\alpha^{-1} \lesssim 3$ ns) or mechanical ($\gamma_\beta^{-1} \lesssim 20$ $\mu$s) lifetime, and we are not aware of any other similarly slow process occurring inside the device.

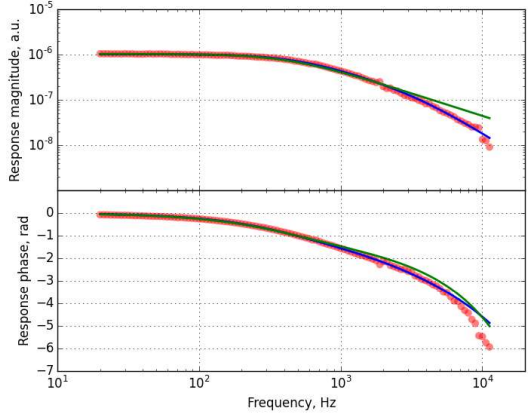

**Figure 3:** Amplitude and phase of the intrinsic linewidth response as a function of the modulation frequency of the circulating optical power. The blue line is the fit to a double exponential (126); for comparison, the green line shows the fit to a simple exponential decay with a time delay, which corresponds to setting $\tau_1 = 0$ in the equation 126.

Finally, we use the measured value of $\tau_0$ to estimate the thermal conductance. First, we need to calculate the heat capacity, for which we can use the known value for the specific heat $c_\nu/T^3 = 8.3 \times 10^{-2}$ J/(mol · K$^4$) [20], which leads to the heat capacity per unit volume $c_V/T^3 = 3 \times 10^3$ J/(K$^4$ · m$^3$). Next, we evaluate the volumes. The cavity has a diameter of $d_{\text{cav}} = 133 \pm 5$ $\mu$m and the length of $l_{\text{cav}} = 70 \mu$m, so its volume is $V_{\text{cav}} = \frac{\pi}{4} d_{\text{cav}}^2 l_{\text{cav}} = (1.0 \pm 0.1) \times 10^{-12}$ m$^3$. The sheaths have the same outer diameter $d_{cav}$ (which is set by the inner diameter of the ferrule), the inner diameter $d_{\text{sh}}$=125 $\mu$m and the length of $l = 3$ mm; this means that the combined volume of two sheaths $V_{\text{sh}} = A_{\text{sh}} l = 2\frac{\pi}{4}(d_{\text{cav}}^2 - d_{\text{sh}}^2)l = (3.5 \div 16) \times 10^{-12}$ m$^3$. The large spread in the volume estimates is due to the uncertainty in the sheath thickness $h_{sh} = (d_{\text{sh}} - d_{\text{cav}})/2 = (1.5 \div 6.5)$ $\mu$m. From the volume estimates we obtain $r_V = 0.06 \div 0.27$ and $C_{\text{sh}}/T^3 = (1.0 \div 4.8) \times 10^{-8}$ J/K$^4$. Using the experimental value for the time constant $\tau_0 = 3.5 \times 10^{-4}$ s we get $\kappa_{\text{sh}}/T^3 \equiv \epsilon = (2.5 \div 7.8) \times 10^{-5}$ W/K$^4$.

We can assess the validity of this result using the theoretical expression for the thermal conductivity of a cylindrical channel[21]:

$$\kappa(T) = \frac{1}{3} C(T) v_{He} d_{\text{ch}} \frac{2-f}{f}, \tag{127}$$

which is applicable when the phonon mean free path is limited by the scattering at the channel boundaries. Here $d_{\text{ch}}$ is the diameter of the channel, and $f$ is the fraction of diffusive phonon scattering events at the channel walls. This equation can be rewritten as

$$\frac{2-f}{f} = \frac{3}{v_{He}d_{ch}}\frac{\kappa(T)}{C(T)} \quad (128)$$

Using equation 125 to express the ratio $C/\kappa$, we find

$$\frac{2-f}{f} = \frac{3}{v_{He}d_{ch}}\frac{l^2}{\tau_0}\left(\frac{4}{\pi^2} + r_V\right) \quad (129)$$

If we approximate $d_{ch}$ by twice the sheath thickness (to account for the much longer longitudinal scattering events) $d_{ch} \approx 2h_{sh} = d_{cav} - d_{sh} = (3 \div 13)$ μm, we will find that the diffusive scattering fraction $f$ is between 3% for the minimal sheath thickness of 1.5 μm and 20% for the maximal sheath thickness of 6.5 μm. These values appear reasonable for the optically smooth glass surfaces of the ferrule and the fiber and typical wavelength of thermal phonons $\lambda_{th} = 2\pi\frac{\hbar v_{He}}{3k_B T} \gtrsim 10$ nm for $T < 0.4$ K.

## SI D  Calibration of the Brownian motion data

Here, we describe the analysis of the thermal fluctuation data (shown in figure 4 of the main text) that is used to extract $\bar{n}_\beta$, the mean phonon number of the acoustic mode. These measurements were taken with the control beam blue-detuned from the cavity resonance by $\omega_\beta$. When the acoustic mode's thermal fluctuations add sidebands to this beam, the blue sideband is strongly suppressed by the cavity lineshape while the red sideband is resonant with the cavity. Beating between the local oscillator and the red sideband of the control beam produces a peak in the photocurrent spectrum which reflects the thermal fluctuations of the mechanical oscillator.

In such a measurement, the power spectral density of the photocurrent, $S_{rr}$ is predicted to be[22]

$$S_{rr} \propto F_{rr} + \frac{\gamma_\beta L_{rr} + (\omega - \omega_\beta)A_{rr}}{(\frac{\gamma_\beta}{2})^2 + (\omega - \omega_\beta)^2} \quad (130)$$

where $F_{rr}$ is the measurement noise floor, $L_{rr}$ is the amplitude of the symmetric (Lorentzian) component, and $A_{rr}$ is the amplitude of the antisymmetric component. When the control beam is detuned from the cavity by $\omega_\beta$ and the laser's classical noise is small, $A_{rr}$ is expected to be negligible and is ignored in the following analysis. The proportionality constant in eq. 130 is also not relevant for the analysis below. Expressions for $F_{rr}$ and $L_{rr}$ are provided in ref. [22]. They depend upon the classical phase noise in the control beam (parameterized by the constant $C_{yy}$[22]), as well as the optical losses and added noise between the cavity and the photodetector.

When $C_{yy} = 0$, the signal-to-noise ratio (SNR) $\frac{L_{rr}}{F_{rr}}$ is given by[22]

$$\frac{L_{rr}}{F_{rr}} = \sigma \kappa_{\alpha,in}\bar{n}_\alpha g_0^{\alpha,\beta\,2}|\chi_{cav}[-\omega_\beta]|^2(\bar{n}_\beta + 1) \quad (131)$$

where $\sigma = 10^{-\frac{\Lambda}{10}}$ is the measurement detection efficiency, $\Lambda$ is the signal-to-noise ratio (SNR) degradation factor in dB, $\kappa_{\alpha,in}$ is the optical cavity input coupling, $\bar{n}_\alpha$ is the mean intracavity photon number due to the control beam, $g_0^{\alpha,\beta}$ is the optomechanical coupling, and $\chi_{cav}[\omega]$ is the cavity susceptibility.

Similar expressions exist for $C_{yy} \neq 0$, [22] but are more cumbersome and so are not displayed here.

Equation 131 (or the equivalent expressions for $C_{yy} \neq 0$) can be inverted to find $\bar{n}_\beta$. In practice, we do this by measuring $\kappa_{\alpha,\text{in}}$, $\bar{n}_\alpha$, $g_0^{\alpha,\beta}$ and $\chi_{\text{cav}}[\omega]$ using standard techniques, determining $L_{\text{rr}}$ and $F_{\text{rr}}$ from fitting the photocurrent spectrum to eq. 130, and measuring the optical losses and added noise that contribute to $\Lambda$. Below, we briefly summarize the individual contributions to $\Lambda$.

- The signal due to the acoustic motion is attenuated by optical loss in the fiber connections, splices, and all of the optical components between the cavity and the EDFA. These losses are measured to sum to $\Lambda_{\text{loss}} = 1.5$ dB.

- The EDFA outputs an amplified beam with additional noise. This added noise depends on the laser wavelength and power. As shown in figure 4a of the main text, two different laser powers $P_{\text{in}}$ were used while acquiring cavity reflection spectra. With $P_{\text{in}} \approx 10(50)\mu\text{W}$, the noise added by the EDFA was measured to be $\Lambda_{\text{EDFA}} = 6(5.1)$ dB.

- The analysis leading to eq.130 assumes that the Lorentzian component arises from beating between the local oscillator and the acoustic sideband, and that the noise floor arises from the beating of the output noise of the EDFA with the local oscillator. However other beams account for 10% - 20% of the power incident on the photodiode; the beating of those beams with the output noise of EDFA increases the noise floor, resulting in $\Lambda_{\text{noise}} = 0.5 - 1.0$ dB.

- Mixer image noise produces additional noise in the photocurrent power spectral density. In order to acquire Brownian motion spectra, the photocurrent signal must be mixed down to within the bandwidth of data acquisition system. This is achieved by mixing the photocurrent with a microwave local oscillator (MLO) with frequency $\omega_{\text{MLO}}$ and analyzing the mixer's output. Due to the symmetric nature of the mixer, the signal at the IF port at a frequency $\omega_{\text{IF}}$ will come from the photocurrent both at $\omega_{\text{MLO}} + \omega_{\text{IF}}$ and $|\omega_{\text{MLO}} - \omega_{\text{IF}}|$. Therefore, assuming a white noise spectrum for the photocurrent signal, the noise at the mixer's IF output is twice the noise at the RF frequency on the mixer's input. To minimize the effect of this added noise, a filter is used to pass the signal band around $\omega_{\text{MLO}} + \omega_{\text{IF}}$, and to partially reject the image band at $|\omega_{\text{MLO}} - \omega_{\text{IF}}|$. As a result, mixer image noise adds only about 1 dB of noise to the photocurrent power spectral density. Hence, $\Lambda_{\text{mix}} = 1.0$ dB.

- In order to measure electronics noise, the noise at the mixer's IF output frequency is measured in the absence of the laser. This noise is much smaller than the added noise produced by the EDFA, and is subtracted from the photocurrent noise floor prior to further calibration.

The SNR degradation $\Lambda$ is given by $\Lambda = \Lambda_{\text{loss}} + \Lambda_{\text{EDFA}} + \Lambda_{\text{noise}} + \Lambda_{\text{mix}}$. Hence, the measurement detection efficiency is 11% to 14% depending on the laser power incident on the EDFA. The photodetector quantum efficiency does not affect the measurement detection efficiency, because the noise on the input of the photodetector is dominated by the classical EDFA noise.

However, to completely describe the SNR degradation of the measurement, we must also consider the laser's classical phase noise. Using a delay line technique[22], the laser's classical phase noise is measured to be $-142.5$ dBc/Hz at frequencies around the mechanical sideband frequency. This corresponds to $C_{yy} \simeq 1$.

There are two ways in which classical laser noise affects the calibration of the Brownian motion data. First, laser phase noise is transformed into amplitude noise through interaction with the optical cavity, and thereby increases the photocurrent noise floor. Based on the analysis in ref. [22] and the measured value of $C_{yy}$, this process can alter $F_{\text{rr}}$ enough to reduce the estimated value of $\bar{n}_\beta$ by $\lesssim 1$.

Second, classical laser noise enters and reflects from the cavity in the same manner as the probe beam used in the OMIA measurements, described in section SI A. As a result, the familiar phenomenon of noise (anti-)squashing occurs; we estimate that this process can alter $L_{\text{rr}}$ enough to increase the estimated value of $\bar{n}_\beta$ by $\lesssim 1$.

Since these two contributions from laser phase noise are small and tend to cancel, they are ignored in the analysis of $\bar{n}_\beta$.